\documentclass[twocolumn,showpacs,aps,pra]{revtex4-1}
\usepackage{amssymb}
\usepackage{amsfonts}
\usepackage{dcolumn}
\usepackage{amsmath}
\usepackage{graphicx}
\usepackage{psfrag,pstricks}
\usepackage{amsmath}
\usepackage{amssymb}

\def\be{\begin{equation}}
\def\ee{\end{equation}}

\begin{document}

\title{Purcell effect with microwave drive: Suppression of qubit relaxation rate}
\author{Eyob A. Sete$^{1}$, Jay M. Gambetta$^{2}$, and Alexander
N. Korotkov$^{1}$}
\affiliation{$^{1}$Department of Electrical Engineering, University
of California, Riverside, California 92521, USA\\
$^{2}$IBM T.J. Watson Research Center, Yorktown Heights, New York
10598, USA}
\date{\today}
\begin{abstract}
We analyze the Purcell relaxation rate of a superconducting qubit
coupled to a resonator, which is coupled to a transmission line and
pumped by an external microwave drive. Considering the typical
regime of the qubit measurement, we focus on the case when the qubit
frequency is significantly detuned from the resonator frequency.
Surprisingly, the Purcell rate decreases when the strength of the
microwave drive is increased. This suppression becomes significant
in the nonlinear regime. In the presence of the microwave drive, the
loss of photons to the transmission line also causes excitation of
the qubit; however, the excitation rate is typically much smaller
than the relaxation rate. Our analysis also applies to a more
general case of a two-level quantum system coupled to a cavity.

\end{abstract}
\maketitle

\section{Introduction}
The spontaneous emission rate of an atom depends on the environment with
which it is coupled. Changing the atom environment substantially
alters the density of states, leading to suppression or enhancement
of the spontaneous emission rate. This phenomenon was first predicted by
Purcell in his seminal work \cite{Pur46}. When the atom is coupled
on-resonance with a cavity, its relaxation rate is enhanced
\cite{Pur46,Goy83} due to increased vacuum fluctuations
at the atom frequency. On the other hand, if the atom is off-resonant from the cavity
frequency, the spontaneous emission rate can be significantly
suppressed \cite{Kle83,Hul85,Jhe87}.

    A very similar effect \cite{Martinis-86} (thus often called
the Purcell effect) occurs in circuit quantum electrodynamics (QED)
systems \cite{Blais-04,Wal04} when a superconducting qubit
(artificial atom) is coupled to a microwave resonator, which in turn
is coupled to a transmission line. The qubit energy relaxation via
the resonator is one of the main processes limiting the qubit
lifetime. As demonstrated experimentally \cite{Hou08}, even coupling
with resonator modes that are far detuned from the qubit frequency
can significantly reduce the qubit lifetime. The Purcell effect is
also one of the limiting factors in achieving a high-fidelity qubit
readout. Several proposals have been put forward to reduce or
eliminate the resonator-induced qubit relaxation rate (Purcell rate)
either by designing a Purcell filter \cite{Red10,Sank-14}, engineering a
Purcell-protected qubit \cite{Gam11}, or using a tunable coupler
\cite{Yin13} that decouples the transmission line from the resonator
during the qubit-resonator interaction, thereby avoiding the Purcell
effect altogether \cite{Set13}.

In this work, we analyze the effect of an external microwave drive
on the qubit relaxation rate caused by the loss of photons to the
resonator environment. It is known \cite{Boi-08,Boi09} that the
external drive can increase the qubit relaxation rate  (and also
cause qubit excitation) due to  the ``dressed dephasing'' effect,
which essentially converts pure dephasing into
photon-number-dependent energy relaxation (the dressed dephasing was observed experimentally \cite{Sli12}). In our analysis we assume the absence of
pure dephasing, so that there is no dressed dephasing, and we can
focus on the question of how the standard Purcell effect changes in the
presence of an additional drive. In spite of the general importance
of this question, we are not aware of any direct discussion
of the Purcell effect in the presence of drive, except for an indirect
analysis in Ref.\ \cite{Boi09}.

\begin{figure}[t]
\includegraphics[width=8cm]{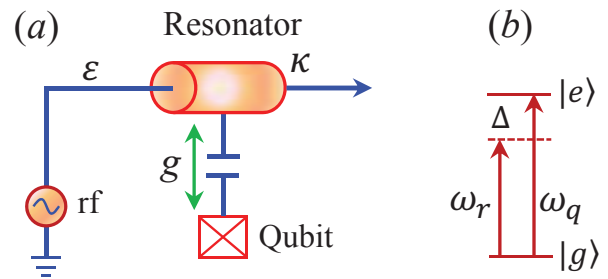}
  \caption{
  (a) Schematic of the considered system with (b) the energy-level
diagram.
  The qubit is off-resonantly ($\Delta=\omega_q-\omega_r$)
coupled to the resonator, with coupling constant $g$. The resonator
decays with the rate $\kappa$, which causes the energy relaxation
(Purcell relaxation) of the qubit. In a circuit QED qubit measurement
setup, a microwave (rf) drive with resonant frequency $\omega_{\bf
r}$ and normalized amplitude $\varepsilon$ is applied to the
resonator. We show that the qubit
relaxation rate decreases with increasing strength of this drive, and that
there exists a relatively weak qubit excitation. }
   \label{fig-model}
\end{figure}

  We consider a superconducting qubit coupled to a resonator, which
can leak photons into a transmission line (Fig.\ \ref{fig-model}).
The resonator is driven on resonance by an external microwave field,
while the qubit is significantly detuned from the resonator
frequency, so that there is no direct effect of the drive on the
qubit. Nevertheless, the presence of microwave photons in the
resonator (with average number $\bar{n}$) may affect the qubit
relaxation rate. Naively, we might expect that the qubit relaxation
rate should scale approximately as $\bar{n}+1$ with increasing drive,
because the effective interaction between the qubit and resonator
scales as $\sqrt{N_{e}}$, where $N_{e}$ is the number of excitations
in the system. However, this is not correct: the Purcell rate does not
increase with $\bar{n}$. It is easy to understand this fact using a
picture of an almost linear interaction between the qubit and
resonator, so that the photons at the resonator frequency do not
affect the evolution of the qubit excitation, which is at the qubit
frequency. Our calculations confirm that the Purcell rate does not
increase with $\bar{n}$ even in the strongly nonlinear regime (when $\bar n \agt n_{\rm crit}$, where $n_{\rm crit}$ is the so-called critical photon number \cite{Blais-04} -- see later).

    In fact, somewhat surprisingly, we find the opposite effect: the
qubit decay rate {\it decreases} with increasing $\bar{n}$. This
follows from an analytic formula, obtained in a simple intuitive approach. In the slightly nonlinear regime (when $\bar n \ll n_{\rm crit}$) this formula  is confirmed
using a formal approach based on a dispersive expansion of the
interaction Hamiltonian; the formula is also confirmed by a numerical
simulation in a wider range of nonlinearity. The suppression of the relaxation rate becomes significant
(it may exceed an order of magnitude) when increasing $\bar{n}$
brings the interaction into the strongly nonlinear regime ($n\agt n_{\rm crit}$). We have also found that
besides the energy relaxation, in the presence of the microwave
drive, the qubit may experience excitation as a result of photon
loss to the resonator environment. In the
moderately nonlinear regime the excitation rate grows with $\bar{n}$, but
remains much smaller than the relaxation rate. The simple analytics and formal
analysis for the relaxation as well as the excitation rate agree well
with the numerical results.

    The paper is organized in the following way. In Sec.\ II we
start with reviewing the standard Purcell effect in the absence of a
microwave drive. Then, in Sec.\ III, we calculate the Purcell rate
in the presence of the drive analytically in two ways: using a
simple approach and using a formal master equation approach. The
analytical results are compared with the numerical results in Sec.\
IV. Finally, Sec.\ V is the conclusion.

\section{Purcell effect without drive}

We begin with a discussion of the Purcell rate calculation
\cite{Haroche-book} by considering a qubit coupled to a resonator in
the absence of a microwave drive. In this case, there are only three (bare)
states involved in the evolution: $|{\bf e}\rangle =
|e\rangle|0\rangle = |e,0\rangle$, $|{\bf 1}\rangle = |g\rangle|1
\rangle = |g,1\rangle$, and $|{\bf g}\rangle = |g\rangle|0\rangle =
|g,0\rangle$, where $|e\rangle$ and $|g\rangle$ denote the qubit
states, while $|0\rangle$ or $|1\rangle$ represents the resonator
state with 0 or 1 photon. We assume that the system is initially in the state $|{\bf
e}\rangle$ or a superposition of $|{\bf e}\rangle$ and $|{\bf
1}\rangle$. Then the qubit-resonator coupling causes coherent
oscillations between the states $|{\bf e}\rangle$ and $|{\bf
1}\rangle$; however, leakage of the photon to the resonator
environment eventually causes irreversible relaxation to the
system ground state $|{\bf g}\rangle$ (note that in some cases the
Purcell relaxation can be considered as a coherent process
\cite{Houck-transfer}).

The qubit-resonator system is described by the usual Jaynes-Cummings
(JC) Hamiltonian ($\hbar=1$)
\begin{align}\label{Ham-JC}
  H=\omega_{\rm r}a^{\dag}a+\frac{1}{2}\omega_{\rm q}\sigma_{z}
  +\textit{g} \, (a^{\dag}\sigma_{-}+ a\sigma_{+}),
\end{align}
where $\omega_{\rm r}$  and $\omega_{\rm q}$ are the resonator and
the qubit frequencies, respectively, and $\textit{g}$ is the
qubit-resonator coupling, assumed to be real for simplicity of
notations. Here $\sigma_{\pm}$ are the qubit raising and lowering
operators,  $\sigma_z=|e\rangle\langle e|- |g\rangle \langle
g|=\sigma_+\sigma_- -\sigma_-\sigma_+$, and $a$ ($a^{\dag}$) is the
annihilation (creation) operator for the resonator field.

    Let us use the rotating frame, which rotates with frequency
$\omega_{\rm r}$ for both the resonator and qubit. Formally, this is
introduced by using the interaction picture, based on separating the
Hamiltonian into two parts, $H=H_{0}+V$, with
\begin{eqnarray}
&&  H_{0} = \omega_{\rm r}a^{\dag}a+
 \frac{1}{2}\omega_{\rm r}\sigma_{z},
 \label{H0}\\
&& V= \frac{\Delta}{2}\sigma_{z}+ \textit{g} \, (a^{\dag}\sigma_{-}+
a\sigma_{+}),
\end{eqnarray}
where $\Delta=\omega_{\rm q}-\omega_{\rm r}$, $H_{0}$ is the free
Hamiltonian, and $V$ is the remaining interaction part. Note
however that in the interaction picture the Hamiltonian does not
depend on time, $V_I =\exp(iH_{0}t)V\exp(-iH_{0}t)=V$, because $V$
commutes with $H_0$. Even though the resonator and qubit operators
now formally depend on time, $a(t)=a(0)\exp(-i\omega_{\rm r }t)$,
$a^\dagger(t)=a(0)\exp(i\omega_{\rm r }t)$, and
$\sigma_{\pm}(t)=\sigma_{\pm}(0)\exp(\pm i\omega_{\rm r }t/2)$, they
always come in pairs, so that the time dependence is canceled out.
This is why we can still use the time-independent operators $a$, $a^\dagger$, and
$\sigma_{\pm}$, which simplifies calculations. This trick is possible because the JC Hamiltonian
(\ref{Ham-JC}) is only the rotating-wave approximation of the actual
physical Hamiltonian.

The evolution of the qubit-resonator system that accounts for the photon
loss from the resonator can be described by a master equation in
Lindblad form \cite{Haroche-book}
\begin{eqnarray}\label{H}
&&  \dot \rho=-i[V,\rho]+\kappa\mathcal{D}[a]\rho,
    \\
\label{LB} && \kappa\mathcal{D}[a]\rho=\kappa(a\rho
a^{\dag}-a^{\dag}a\rho/2-\rho a^{\dag} a/2),
 \end{eqnarray}
where $\kappa$ is the resonator decay rate. Note that as discussed
above, all operators here (except the density matrix $\rho$) do not
depend on time, even though we are using the interaction
picture.
Also note that in this master equation we assumed that coupling between the resonator and the bath (transmission line) is frequency-independent; this assumption is no longer valid in the case of the Purcell filter \cite{Red10,Sank-14}, when $\kappa$ becomes frequency-dependent.

Since we are only interested in quantifying the qubit relaxation
through the resonator, we do not take into account the intrinsic
qubit relaxation and pure dephasing. Using Eq.\ \eqref{H}, one
easily obtains the equations for the density matrix elements in the
single-excitation subspace (in the bare-state basis of $|{\bf
e}\rangle$ and $|{\bf 1}\rangle$), which are decoupled from the
elements containing state $|{\bf g}\rangle$,
\begin{equation}\label{M}
 \left(
                \begin{array}{c}
                  \dot\rho_{\bf e e} \\
                  \dot\rho_{\bf 1 1} \\
                  \dot \rho_{\bf e 1}^{-}\\
                 \dot\rho_{\bf e 1}^{+}\\
                \end{array}
              \right)=
  \left(
    \begin{array}{cccc}
      0  & 0 & i\textit{g}&0 \\
     0 & -\kappa & -i\textit{g}&0 \\
      2i\textit{g}& -2i\textit{g}& -\frac{\kappa}{2}&-i\Delta \\
0& 0&-i\Delta& -\frac{\kappa}{2}\\
    \end{array}
  \right)\left(
           \begin{array}{c}
                  \rho_{\bf e e} \\
                  \rho_{\bf 1 1} \\
                   \rho_{\bf e 1}^{-}\\
                 \rho_{\bf e 1}^{+}\\
                \end{array}
                      \right),
\end{equation}
where $\rho_{\bf e 1}^{\pm}=\rho_{\bf e 1}\pm\rho_{\bf 1 e}$. Here
the bare-state populations $\rho_{\bf e e}$ and $\rho_{\bf 1 1}$ are
coupled to the coherences $\rho_{\bf e 1}^{\pm}$ via terms proportional
to $g$, while the damping term proportional to
$\kappa$ contributes to the decay of $\rho_{\bf 1 1}$, but does not
affect $\rho_{\bf e e}$. Note that the coherences $\rho_{\bf
e1}^{\pm}$ decay with the rate $\kappa/2$. The population of the
state $|{\bf g}\rangle$ obviously increases as $\dot\rho_{\bf
gg}=\kappa \rho_{\bf 11}$.
    Equation (\ref{M}) can be used in numerical
as well as in analytical calculations. In particular, the eigenvalues of
the evolution can be found analytically from the corresponding
quartic equation, which in this case has a relatively simple
solution. Below we will obtain these eigenvalues in a different way.

    Instead of using the density matrix language for the
description of the Purcell effect, it is also possible to use the
simpler language of wavefunctions, even in the presence of the decay
$\kappa$. Physically, this is because in the single-excitation
subspace unraveling of the Lindblad equation (see, e.g.,
\cite{Kor-13}) corresponds to only one scenario with no relaxation,
and therefore the wavefunction evolution is non-stochastic.
  More formally, we can rewrite the master equation
\eqref{H} as \cite{Pie07,Car93} $\dot \rho=-i[H_{\rm eff},\rho]+\kappa
a\rho a^{\dag}$, where $H_{\rm eff}=V-i\kappa a^{\dag}a/2$ is an
effective non-Hermitian Hamiltonian. The term $\kappa a\rho
a^{\dag}$ can be neglected because in the single-excitation subspace it produces a contribution only from higher-excitation subspaces, which are not populated. Therefore, in the single-excitation
subspace we have $\dot \rho=-i[H_{\rm eff},\rho]$ or equivalently $|\dot
\psi \rangle=-iH_{\rm eff}|\psi\rangle$, which describes the
evolution of the {\it decaying} wavefunction $|\psi (t)\rangle
=\alpha(t) |{\bf e}\rangle + \beta(t) |{\bf 1}\rangle$:
    \begin{eqnarray}
&& \dot{\alpha} = -i \frac{\Delta}{2}\alpha -i\textit{g} \beta ,
    \\
&& \dot{\beta} = i \frac{\Delta}{2}\beta -i\textit{g} \alpha
-\frac{\kappa}{2}\beta .
    \end{eqnarray}
 (Another way to derive these equations is by considering
only the no-relaxation scenario when unraveling the evolution
\cite{Kor-13}.)
 These are the usual equations for a
two-level system, but with complex energy $-\Delta/2-i\kappa/2$ of
the bare state $|{\bf 1}\rangle$. Using the standard diagonalization
procedure, we can find two eigenstates with energies
$\tilde{E}=-i\kappa/4 \pm \sqrt{-\kappa^2/16+\textit{g}^2
+\Delta^2/4 +i\kappa\Delta/4}$, which can be written as
    \begin{equation}
    \tilde{E}_{ e} =\frac{\Omega}{2} -
    i\frac{\Gamma}{2}, \quad
 \tilde{E}_{ 1} =-\frac{\Omega}{2} -
    i\frac{\kappa-\Gamma}{2}, \label{tilde-E}
\end{equation}
with
\begin{eqnarray}
&&    \Gamma =
    \frac{\kappa}{2}-\frac{\sqrt{2}}{2}\sqrt{-A+\sqrt{A^2+(\kappa\Delta)^2}},
    \label{Gamma-exact}\\
 && \Omega = \frac{\sqrt{2}}{2}\sqrt{A+\sqrt{A^2+(\kappa\Delta)^2}}\,
 {\rm sgn}(\Delta) ,
    \label{Omega-exact}\\
&& A=\Delta^2 +4\textit{g}^2 -\kappa^2/4 .
    \end{eqnarray}
Here $\tilde E_{ e}$ is the complex energy of the eigenstate ({\it
``dressed'' state, which includes decay}), corresponding to the
excited qubit, while $\tilde E_{ 1}$ corresponds to the dressed
resonator photon (notice the sign of $\Delta$ in the formula for
$\Omega$).
Since ${\rm Im}(\tilde E_{ e})=-\Gamma/2$, the population of the qubit dressed state decays with the rate $\Gamma$.
 Therefore $\Gamma$ is the qubit relaxation rate, i.e.,
the Purcell rate, while the photon relaxation rate is
$\kappa-\Gamma$. If the initial state is not one of these
eigenstates, then the evolution also includes oscillations with
beating frequency $\Omega$, decaying with the rate $-{\rm
Im}(\tilde{E}_{ e}+\tilde{E}_{ 1})=\kappa/2$. We have checked that
the same rates can be obtained by diagonalizing the evolution matrix
in Eq.\ \eqref{M}, which has four eigenvalues: $-\Gamma$, $-(\kappa
-\Gamma)$, and $-\kappa/2\pm i\Omega$, as expected from Eq.\
(\ref{tilde-E}). Note that for small $\kappa$ the frequency $|\Omega
|$ is close to the usual Rabi frequency $\sqrt{\Delta^2+4\textit{g}^2}$,
but large $\kappa$ may change it significantly.

    The eigenstates (dressed states with decay)
$|{\bf \tilde{e}}\rangle =
    \tilde\alpha_{ e}|{ \bf e}\rangle +\tilde\beta_{
e}|\bf{1}\rangle$ and $|{\bf \tilde{1}}\rangle =
    \tilde\alpha_{ 1}|{ \bf e}\rangle +\tilde\beta_{
1}|\bf{1}\rangle$  corresponding to the energies $\tilde E_{ e}$ and
$\tilde E_{ 1}$ can be found in the standard way, via the ratio
$\tilde\beta_{{ e},{ 1}}/\tilde\alpha_{{ e},{ 1}}=(\tilde{E}_{{ e},{
1}}-\Delta/2)/\textit{g}$ and the normalization condition. However,
note that these eigenstates are not mutually orthogonal, so
finding the expansion of an initial state in the eigenbasis, $|\psi_{\rm
in}\rangle =\tilde{c}_{ e} |{\bf \tilde{e}}\rangle +\tilde{c}_{ 1}
|{\bf \tilde{1}}\rangle$, is somewhat more involved:
$\tilde{c}_1=\langle {\bf \tilde{e}}_\perp|\psi_{\rm in}\rangle/
\langle {\bf \tilde{e}}_\perp|{\bf \tilde{1}}\rangle$, where $|{\bf
\tilde{e}}_\perp\rangle$ is the vector orthogonal to $|{\bf
\tilde{e}}\rangle$ (similarly for $\tilde{c}_e$). The evolution in the single-excitation subspace
is then $|\psi (t)\rangle = \tilde{c}_{ e} e^{-i\Omega
t/2}e^{-\Gamma t/2} |{\bf \tilde{e}}\rangle +\tilde{c}_{ 1}
e^{i\Omega t/2}e^{-(\kappa- \Gamma) t/2} |{\bf \tilde{1}}\rangle$,
and it is easy to find the density matrix elements; for example the
bare-state qubit occupation is $\rho_{\bf ee} (t) = |\tilde c_e
\tilde \alpha_{ e} e^{-i\Omega t/2}e^{-\Gamma t/2}  +\tilde c_1
\tilde\alpha_{ 1} e^{i\Omega t/2}e^{-(\kappa- \Gamma) t/2} |^2$.

Now let us consider several special cases starting with the case
usually discussed for optical systems (e.g., \cite{Haroche-book}),
in which the resonator damping rate is much larger than the
coupling: $\kappa\gg |\textit{g}|$, while $\Delta$ is arbitrary.
Expanding Eqs.\ (\ref{Gamma-exact}) and (\ref{Omega-exact}) in
$\textit{g}/\kappa$ and keeping the first two leading terms produces
\begin{eqnarray}
\label{prc} && \Gamma = \frac{\kappa
\textit{g}^2}{\Delta^2+(\kappa/2)^2},
    \\
&&  \Omega =
\Delta\left(1+\frac{2\textit{g}^2}{\Delta^2+(\kappa/2)^2}\right) .
\qquad
\end{eqnarray}
   This formula for the Purcell rate $\Gamma$ can be interpreted as
Fermi's golden rule for the transition from the state $|{\bf
e}\rangle$ to the relaxation-widened state $|{\bf 1}\rangle$ with
width $\kappa$. Note that at resonance, $\Delta=0$, the Purcell rate
(\ref{prc}) reduces to $\Gamma = 4g^2/\kappa$, while in the case
$|\Delta |\gg \kappa$ it becomes $\Gamma = \kappa g^2/\Delta^2$;
also note that $\Gamma \ll \kappa$ because of the assumption
$|\textit{g}|\ll \kappa$. If the evolution starts with the bare
state $|{\bf e}\rangle$, then the probability to find the qubit in
the excited state decays as
\begin{align}\label{raa}
  \rho_{\bf e e}(t)& \approx e^{-\Gamma t}\left[1+\frac{8\textit{g}^2(\kappa^2-4\Delta^2)}{(\kappa^2+4\Delta^2)^2}\right]-\frac{8\textit{g}^2 e^{-\kappa t/2}}{(\kappa^2+4\Delta^2)^2}\notag\\
&\times \left[4\kappa|\Delta| \sin (|\Omega| t)+(\kappa^2-4\Delta^2)\cos
(\Omega t)\right].
\end{align}
 In this formula the oscillation amplitude is always small because of
the assumption $|\textit{g}| \ll \kappa$ (the amplitude of the
neglected term $e^{-(\kappa -\Gamma)t}$ is even smaller), so that in
the leading order $\rho_{\bf e e}(t) \approx e^{-\Gamma t}$. We have
obtained Eq.\ (\ref{raa}) from the density matrix evolution
(\ref{M})
by using the method of Laplace transform, in which  $\rho (t) \rightarrow \rho (s)$ and
$\dot \rho(t)\rightarrow s\rho (s)-\rho (t=0)$. The solution is then obtained by taking the inverse Laplace transform of $\rho (s)$.

   In circuit QED experiments the typical regime
is different from what is usually considered in optics. Most
importantly, instead of the above assumption $\kappa \gg
|\textit{g}|$, in the circuit QED systems the relation is usually the
opposite: $\kappa \alt |\textit{g}|$ or even $\kappa \ll
|\textit{g}|$. Therefore, the result (\ref{prc}) for the Purcell rate
is inapplicable in the way we derived it, but we can still use the
exact formula (\ref{Gamma-exact}).
  [Note that Eq.\  (\ref{prc}) can actually be derived from Eq.\
(\ref{Gamma-exact}) using a weaker assumption: $|\textit{g}|\ll \max
(\kappa, |\Delta|)$.]
     The formula for the Purcell rate that is most widely used in
the circuit QED is the strong-dispersive-regime formula
\cite{Blais-04}
    \begin{equation}\label{gcd}
\Gamma_{d}=\frac{\kappa \textit{g}^2}{\Delta^2},
    \end{equation}
which can be derived from Eq.\ (\ref{Gamma-exact}) by assuming
$|\Delta|\gg \max (|\textit{g}|,\kappa)$, keeping the relation between
$|\textit{g}|$ and $\kappa$ arbitrary. This formula has a simple
interpretation: the fraction $(\textit{g}/\Delta)^2$ of the dressed
qubit state (i.e.\ the eigenstate) is in the form of the resonator
photon, and this part decays with the rate $\kappa$.

    Now let us derive the expression for the Purcell rate using the
weaker assumption $\kappa \ll \sqrt{\Delta^2 +4\textit{g}^2}$, which
is usually well-satisfied in the circuit QED experiments. The
detuning $|\Delta|$ may be much larger or comparable to the coupling
$|\textit{g}|$.  In this case it is physically natural to use the
basis of the single-excitation eigenstates (dressed states without
decay) $\overline{|e,0\rangle}$ and $\overline{|g,1\rangle}$ besides
the basis of the bare states $|{\bf e}\rangle =|e,0\rangle$ and
$|{\bf 1}\rangle =|g,1\rangle$. (Note the difference from the
eigenbasis $\{ |{\bf \tilde{e}}\rangle, |{\bf \tilde{1}}\rangle\}$,
which includes decay.) If the initial state is
$\overline{|e,0\rangle}$, then the system would remain in this state
if $\kappa=0$. However, non-zero (but still small) $\kappa$ causes
the rare jumps $|g,1\rangle \rightarrow |g,0\rangle$. Therefore, the
rate of jumps should be proportional to the occupation of the bare
state $|g,1\rangle$ and the Purcell rate can be approximated as
\begin{equation}\label{ga}
  \Gamma_{P}=\kappa |\langle g,1|\overline{e,0\rangle}|^2=
  \frac{\kappa}{2}\left(1-\frac{\Delta}{\sqrt{\Delta^2+4g^2}}\right),
\end{equation}
where we used the exact formula for the overlap between the states
$\overline{|e,0\rangle}$ and $|g,1\rangle$. Numerical comparison
between this formula and exact result (\ref{Gamma-exact}) shows that
it is a very good approximation for small $\kappa$; in particular,
the relative error is less than $0.25\kappa^2/(\Delta^2+4g^2)$ for
$\kappa/g<4$. We will be using Eq.\ \eqref{ga} in the following
sections. It is easy to see that in the regime $|\Delta| \gg
|\textit{g}|$ the rate $\Gamma_P$ reduces to $\Gamma_{d}$ in Eq.\
\eqref{gcd}.

    If the evolution starts with the bare state
$|e,0\rangle$, then in the dispersive regime, $|\Delta|\gg
|\textit{g}|\agt \kappa$, its population evolves as
    \begin{align}\label{rab}
  \rho_{\bf e e}(t) \approx e^{-\Gamma_{P} t}
  \left(1-\frac{2\textit{g}^2}{\Delta^2}\right)+
  \frac{2\textit{g}^2}{\Delta^2} \cos (\Omega t) \, e^{-\kappa t/2},
\end{align}
where $|\Omega| \approx \sqrt{\Delta^2 + 4\textit{g}^2}$. The
oscillation amplitude is small, but still noticeable for moderate
values of $\Delta/\textit{g}$. In contrast, if the evolution starts
with the eigenstate $\overline{|e,0\rangle}$, then the population of
this eigenstate decays almost without oscillation,
    \be
    \overline{\rho}_{\bf ee} \approx e^{-\Gamma_P t}
    \left(1+\frac{\textit{g}^2\kappa^2}{2\Delta^4}\right)
    -\frac{\textit{g}^2\kappa^2}{2\Delta^4}\cos(\Omega t)e^{-\kappa t/2}
    \approx e^{-\Gamma_P t} .
    \label{rho-eigen-disp}\ee
The small remaining oscillation amplitude
$\textit{g}^2\kappa^2/2\Delta^4$ is due to the
difference between $\overline{|e,0\rangle}$ and $|{\bf
\tilde{e}}\rangle$, while the amplitude $2\textit{g}^2/\Delta^2$ in
Eq.\ (\ref{rab}) is due to the much larger difference between
$|e,0\rangle$ and $\overline{|e,0\rangle}$ (the factor of 2 comes
from two conversions: from the bare basis to the eigenbasis and then back).

\section{Purcell effect with microwave drive}

So far we have only considered one excitation in the system. This
situation is relevant to the qubit decay during coherent operations
in circuit QED systems. However, the qubit measurement
\cite{Blais-04,Joh12,Ris12,X-mon13} requires adding a microwave
drive in resonance (or close to resonance) with the resonator
frequency -- see Fig.\ \ref{fig-model}. This is the main motivation
here to analyze qubit relaxation through the resonator in the
presence of additional excitations in the system. To this end, we
consider a dispersive qubit-resonator interaction (but not
necessarily strongly dispersive, say $|\Delta/\textit{g}|\geq 5$)
with an external microwave drive. The Hamiltonian for the
qubit-resonator system with a coherent microwave drive is given by
\begin{align}\label{Ham}\
  H&=\omega_{\rm r}a^{\dag}a+\frac{1}{2}\omega_{\rm q}\sigma_{z}+\textit{g}\,
  (a^{\dag}\sigma_{-}+ a\sigma_{+})\notag\\
&+\varepsilon(a e^{i\omega_{\rm d}t}+a^{\dag} e^{-i\omega_{\rm
d}t}),
  \end{align}
 where $\omega_{\rm d}$ is the drive
frequency and $\varepsilon$, assumed to be real and constant, is the
normalized amplitude of the microwave drive.

   Following the same line of reasoning as was used to derive
Eq.\ \eqref{H}, we introduce the frame rotating at $\omega_{\rm r}$
via the free Hamiltonian (\ref{H0}), so that the interaction part of the
Hamiltonian in this frame has the form
\begin{eqnarray}
&& H_{I}=\frac{\Delta}{2}\sigma_{z}+\textit{g}\,
(a^{\dag}\sigma_{-}+ a\sigma_{+})
    \nonumber \\
 && \hspace{0.8cm}
+\varepsilon \left( e^{i(\omega_{\rm d}-\omega_{\rm r})t}a
+e^{-i(\omega_{\rm d}-\omega_{\rm r})t}a^{\dag}\right),
  \label{ham}\end{eqnarray}
where all operators are time-independent. For simplicity we assume
the drive to be exactly on resonance with the resonator frequency, $\omega_{\rm d}=\omega_{\rm
r}$, though this assumption is not critical for our analysis.
   The master equation, including the loss of photons through
the resonator [see Eq.\ (\ref{LB})], is given by
\begin{equation}\label{mast}
  \dot \rho=-i[H_{I},\rho]+\kappa\mathcal{D}[a]\rho.
\end{equation}

\begin{figure}[t]
\includegraphics[width=8cm]{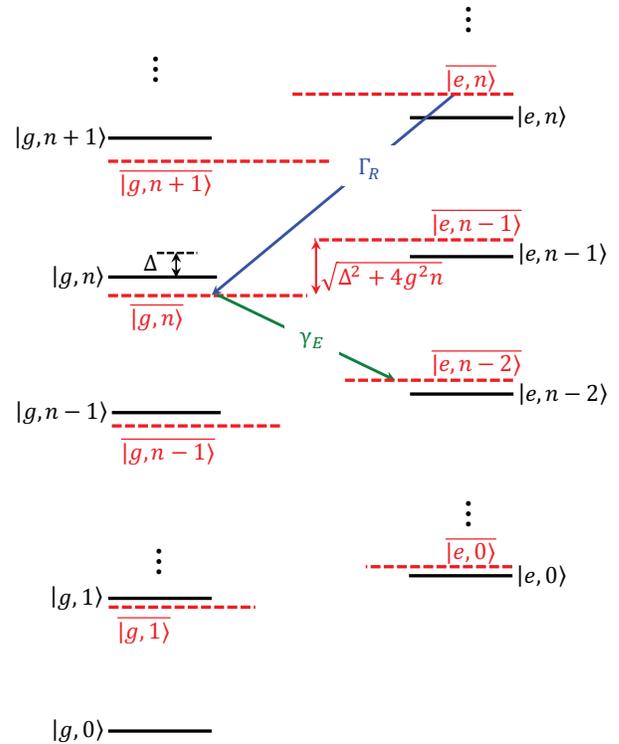}
  \caption{The Jaynes-Cummings ladder in the bare-state basis
(black solid lines) and the eigenstate basis (red dashed lines).
  Here transitions ("jumps") from the right set to the left
set of eigenstates correspond to qubit relaxation with rate
$\Gamma_R$, while transitions from the left to the right set of
eigenstates give rise to qubit excitation with rate $\gamma_E$.}
 \label{fig-JC}
\end{figure}

We focus on the experimentally important regime of sufficiently
small resonator damping rate, $\kappa \ll |\Delta|$. More precisely, we assume
$\kappa \ll \sqrt{\Delta^2 +4g^2(\bar{n}+1)}$, where $\bar{n}$ is
the average number of photons in the resonator, induced by the
drive. In this case the Jaynes-Cummings ladder of levels
\cite{JC-ladder} (Fig.\ \ref{fig-JC}) is affected by an
interaction (between $|e,n\rangle$ and $|g,n+1\rangle$) with the strength $\sqrt{N_e} \, g$,  but the effect of
$\kappa$ is relatively small. Here $N_e$ is the total number of
excitations, $N_e=n$ for the bare state $|g,n\rangle$ and $N_e=n+1$
for the state $|e,n\rangle$. Therefore, it is natural to introduce
the basis of the pairwise eigenstates represented by red dashed
lines in Fig.\ \ref{fig-JC} and denoted by the overline:
    \begin{eqnarray}
&&
\overline{|e,n\rangle}=\cos\theta_{n+1}|e,n\rangle-\sin\theta_{n+1}
   |g,n+1\rangle ,
   \label{en-bar}\\
&&   \overline{|g,n\rangle}=\cos\theta_n|g,n\rangle+
   \sin\theta_n|e,n-1\rangle
    \label{gn-bar},
    \\
    && \hspace{0.0cm} \tan (2\theta_{n})=2g\sqrt{n}/\Delta .
    \label{tan-2theta}
  \end{eqnarray}
The level splitting between the eigenstates is
    \begin{eqnarray}
   \overline{E}_{e,n-1}- \overline{E}_{g,n} =&
    \sqrt{\Delta^2+4n\textit{g}^2} \,\, {\rm sgn} (\Delta)
    \nonumber \\
    = & \Delta
    \sqrt{1+4N_e\textit{g}^2/\Delta^2},
    \label{E-eigen}\end{eqnarray}
and in the rotating frame, which we use, both energies are symmetric about zero,
$\overline{E}_{g,n}=-\overline{E}_{e,n-1}$.  Note that the level splitting changes significantly when the number of photons $n$ becomes comparable with the so-called critical photon number \cite{Blais-04},
    \be
    n_{\rm crit}= \frac{\Delta^2}{4g^2} .
    \label{n-crit}\ee

The use of eigenstates as logical states in quantum computing
applications is more natural than the use of the bare states
\cite{Galiautdinov-12}, and in most practical cases the dynamics is
sufficiently adiabatic to be naturally represented in the eigenbasis
\cite{Set13}. Therefore, when we will consider the qubit being
initially in the excited state, we will actually assume that the
initial state is the (coherent-state) superposition of {\it
eigenstates} (\ref{en-bar}) corresponding to the excited state of
the qubit (right set of red dashed lines in Fig.\ \ref{fig-JC}).
Accordingly, the relaxation process corresponds to increasing
occupation of the $|g\rangle$-eigenstates (\ref{gn-bar}) (left set
of red dashed lines in Fig.\ \ref{fig-JC}).
    The use of eigenstates allows us to avoid oscillations in
the evolution, caused by the difference between the bases [see,
e.g., Eqs.\ (\ref{rab}) and (\ref{rho-eigen-disp}) for the
single-excitation case].

In the following, we present two ways to derive an approximate analytical formula for the Purcell rate in the presence of a microwave drive.
 We first use a simple intuitive approach applicable for an arbitrary nonlinearity (arbitrary ratio $\bar n/n_{\rm crit}$) and then use a formal perturbative approach applicable in the slightly nonlinear regime ($\bar n \ll n_{\rm crit}$).  The results of the two approaches are shown to coincide in the validity range of the formal approach.

\subsection{Purcell rate with drive: Simple approach}

The damping of the resonator state is described by the Lindblad form
\eqref{LB}. We can think of this process by unraveling it into
``jump'' and ``no jump'' scenarios (see, e.g., \cite{Kor-13}), so
that the first term in Eq.\ (\ref{LB}) describes random jumps due to
the action of the operator $a$, which occur with the rate $\kappa
{\rm Tr}(a^\dagger a \rho)=\kappa\bar{n}$, while the remaining two
terms describe the state evolution due to absence of jumps. Without
the qubit the driven resonator would reach a steady coherent
state, for which the jump produces no change (because the coherent
state is an eigenstate of the operator $a$), while the photon-number
decay due to the no-jump evolution \cite{Kor-13} is compensated by
the drive. In the presence of the qubit there are two sets of
eigenstates (Fig.\ \ref{fig-JC}, we will refer to them as two
ladders) and within each ladder the process is approximately the
same as without the qubit (up to a factor, accounting for the
overall population of each ladder). However, the jumps {\it between}
the ladders lead to Purcell-effect relaxation (or excitation).

\subsubsection*{Qubit relaxation rate}

    If the system is in the
eigenstate $\overline{|e,n\rangle}$ (see Fig.\ \ref{fig-JC}), then
the (random) action of the operator $a$ gives the unnormalized state
$a \overline{|e,n\rangle}$. This state is a superposition of bare
states $|e,n-1\rangle$ and $|g,n\rangle$. Expanding the state $a
\overline{|e,n\rangle}$ in the eigenbasis, we see that most of it
contributes to $\overline{|e, n-1\rangle}$, which belongs to the set
of excited eigenstates. However, there is a non-zero overlap between
$a \overline{|e,n\rangle}$ and $\overline{|g,n\rangle}$, which leads
to the qubit energy relaxation (transition from the right to left
ladder of eigenstates in Fig.\ \ref{fig-JC}). The rate of this
relaxation process \cite{Koch-07} (Purcell rate from the state
$\overline{|e,n\rangle}$) is therefore
    \be
 \Gamma_{R}(n)=\kappa|\overline{\langle g,n|}a\overline{|e,n\rangle}|^2,
    \label{Gamma-R-n}\ee
where the subscript $R$ means relaxation. Note that
$\Gamma_R(0)=\Gamma_P$ [see Eq.\ (\ref{ga})]. Using explicit
expressions (\ref{en-bar}) and (\ref{gn-bar}) for the eigenstates,
we find
   \be
  \Gamma_{R}(n)=
\kappa (\sqrt{n+1}\sin\theta_{n+1}\cos\theta_n
-\sqrt{n}\sin\theta_n\cos\theta_{n+1})^2,
    \label{Gamma-R-n-2}\ee
where $\theta_n$ is given by Eq.\ (\ref{tan-2theta}).

It is important that there is a significant energy shift ($\simeq
\sqrt{\Delta^2 +4 \textit{g}^2 \bar{n}}$) between the two ladders of
eigenstates, so that the contribution to $\overline{|g,n\rangle}$
due to the next jump from $\overline{|e,n\rangle}$ is incoherent
with the previous jump contribution (it brings a different random
phase), which allows us to characterize the relaxation process by a
rate $\Gamma_R$. However, when different eigenstates
$\overline{|e,n\rangle}$ are populated, then the jumps from these
states occur in a mutually coherent way (the operator $a$ acts on
the superposition), resulting in a mutual coherence within the set
of eigenstates in the left ladder in Fig.\ \ref{fig-JC}.
Nevertheless, this does not affect the jumps between the left and
right ladders of states, and therefore the rate $\Gamma_R(n)$ can be
simply averaged over the population $P(n)$ of the states
$\overline{|e,n\rangle}$,
    \be
  \Gamma_{R} =\left [ \sum_{n=0}^{\infty} P(n) \, \Gamma_{R}(n) \right] / \sum_{n=0}^{\infty} P(n).
    \label{Gamma-R}
    \ee
A natural assumption is that $P(n)$ is close to the coherent-state distribution (the normalization is not important),
    \be
 P(n)= e^{-\bar{n}} \bar{n}^n /n! \, .
    \label{coh-distr}
    \ee
Actually, this coherent-state assumption becomes invalid at sufficiently large  $\bar n$, when non-linearities (like squeezing) become significant; however, this is not important because at $\bar n \gg 1$ the averaging (\ref{Gamma-R}) is not really needed as long as the spread of occupied values of $n$ is much smaller than $\bar n$.
 For $\bar n$, it is natural to use the value in the absence of
qubit, $\bar n\approx 4|\varepsilon|^2/\kappa^2$, or a more accurate
value, which accounts for the shift of the resonator frequency by
the qubit -- see Eq.\ (\ref{mn}) later. Note that here $\bar{n}$ is
the average number of photons in the ladder of ``excited''
eigenstates (it would be better to use a different notation, $\bar n_e$,
for this meaning -- see the next subsection, but we will use $\bar
n$ here for brevity).

\begin{figure}[t]
\includegraphics[width=8cm]{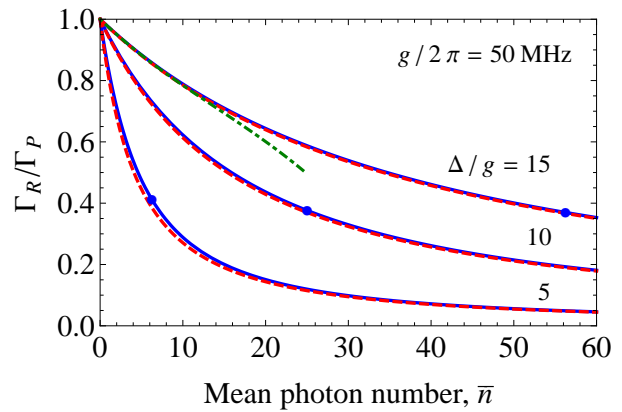}
  \caption{The Purcell relaxation rate $\Gamma_{R}$ normalized
by the no-drive value $\Gamma_{P}$, as a function of the mean photon
number $\bar n$ induced by the drive for several values of
normalized detuning $\Delta/g$: 15, 10, and 5. The dashed red lines
show $\Gamma_R(\bar n)$ calculated using Eq.\ (\ref{Gamma-R-n-2}),
the blue solid lines show $\Gamma_R$ averaged over the coherent
state distribution using Eq.\ (\ref{Gamma-R}). For the upper curve
we also show the truncated expansion [up to $\lambda^8$, Eq.\ (\ref{Gamma-R-exp})] by the
green dot-dashed line. The blue dots on the lines indicate $n_{\rm
crit}$.}
 \label{fig-Gamma}
\end{figure}

The blue solid lines in Fig.\ \ref{fig-Gamma} show the $\bar n$-dependence of the Purcell rate $\Gamma_{R}$ (normalized by the no-drive value
$\Gamma_{P}$) for several values of the normalized detuning
$\Delta/\textit{g}$. As we see, the Purcell rate decreases with increasing $\bar n$, and the suppression (compared with the no-drive case) can be strong at large $\bar n$. The dashed red lines show $\Gamma_{R} (\bar n)$ calculated using Eq.\ (\ref{Gamma-R-n-2}), in which a non-integer $\bar n$ is introduced in the natural way via a non-integer  $\bar n$  in Eq.\ (\ref{tan-2theta}). As expected, the dashed and solid lines almost coincide at $\bar n \gg 10$, while there is a noticeable difference between them for $\bar n \alt 10$ when $\Delta /\textit{g}$ is not very large.

In the experimentally interesting regime when $\Delta \gg g$, the
rate $\Gamma_{R}$ can be expanded in powers of
    \be
\lambda = g/\Delta.
    \ee
(We will need this expansion for comparison with the results of the
formal approach in the next subsection.)
Expanding Eq.\ (\ref{Gamma-R-n-2}) in powers of $\lambda$, we obtain $\Gamma_{R}(n)=\kappa \lambda^2 [ 1-3\lambda^2(2 n+1) +\lambda^4(31 n^2+31 n+10)     -\lambda^6(150 n^3+225 n^2+145 n+35)+...]$.
   The averaging (\ref{Gamma-R}) then    gives $\overline{n^2}=\bar n^2+\bar n$, $\overline{n^3}=\bar n^3+3\bar n^2+\bar n$, etc., thus resulting in a series
\begin{align}\label{Gamma-R-exp}
  \Gamma_{R}&=\kappa \lambda^2\big[1-3\lambda^2(2\bar n+1)
  +\lambda^4(31\bar n^2+62\bar n+10)\notag\\
    &-\lambda^6(150\bar n^3+675\bar n^2+520\bar n+35)+...\big].
\end{align}
This truncated series is shown by the green dot-dashed line in Fig.\ \ref{fig-Gamma}.

The first term in the expansion (\ref{Gamma-R-exp}) gives the standard Purcell rate
$\Gamma_{d}=\kappa \textit{g}^2/\Delta^2$ in the strong dispersive regime
($\lambda\ll 1$) -- see Eq.\ (\ref{gcd}). The negative sign of the
second term means that the Purcell rate decreases with increasing
number of photons in the resonator.  In the leading order of the
correction (assuming $1\ll \bar{n} \ll n_{\rm crit}$) Eq.\
(\ref{Gamma-R-exp}) becomes
    \be
    \Gamma_R \approx \Gamma_d \left( 1-\frac{3}{2}\,
    \frac{\bar{n}}{n_{\rm crit}}\right),
    \label{Gamma-R-approx}\ee
where $n_{\rm crit}$ is given by Eq.\ (\ref{n-crit}).
Note that Eq.\ (\ref{Gamma-R-approx}) gives a slightly inaccurate value for $\bar n \alt 1$ since at $\bar{n}=0$ the expansion (\ref{Gamma-R-exp}) gives
$\Gamma_R(0)=\Gamma_P$ [see Eq.\ (\ref{ga})]; however, the difference between  $\Gamma_d$ with $\Gamma_P$  is small when $|\textit{g}/\Delta|\ll 1$.

    When $\bar{n}\gg 1$, we do not need the summation in
Eq.\ (\ref{Gamma-R}) and can use approximation $\Gamma_R \approx
\Gamma_R (\bar{n})$ if the effective spread of $n$ values is much smaller than $\bar n$, i.e., $\sqrt{\, \overline{n^2}-\bar n^2} \ll \bar n$ (therefore, moderate nonlinear effects still allow this simplification).
 In this case $|\theta_{\bar n+1}-\theta_{\bar n}|\ll 1$ in Eq.\
(\ref{Gamma-R-n-2}), and using the first-order expansion of this
equation we obtain the approximation
    \be
    \Gamma_R \approx \frac{\kappa \textit{g}^2}{4\Delta^2}
    \left(\frac{1}{1+\bar n/n_{\rm crit}}
    +\frac{1}{\sqrt{1+\bar n/n_{\rm crit}}}\right)^2 ,
    \label{Gamma-large-n}\ee
which is valid when $\bar{n}\gg 1$ with arbitrary ratio $\bar n/n_{\rm crit}$. (This is in contrast to the truncated perturbative expansion in powers of $\lambda$, which works well only for $\bar n/n_{\rm crit} \ll 1$.)

\begin{figure}[t]
\includegraphics[width=7.8cm]{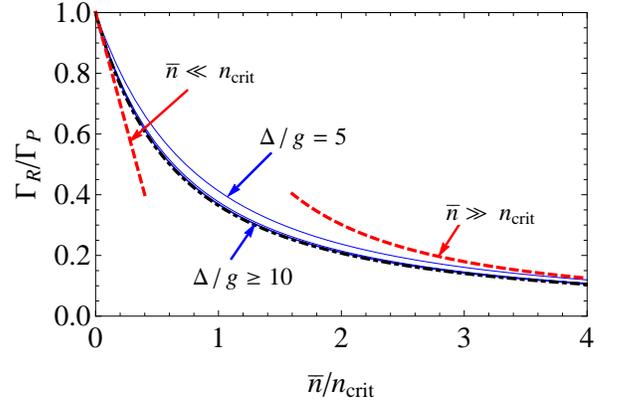}
  \caption{The Purcell relaxation rate $\Gamma_{R}$ [Eq.\
(\ref{Gamma-R})] normalized by the no-drive value $\Gamma_{P}$ [Eq.\
(\ref{ga})] versus $\bar n/n_{\rm crit}$ for several values of
$\Delta/g$: 5, 10, 15, and 20 (blue solid lines). The curves for
$\Delta/g\geq 10$ are practically indistinguishable from each other.
The dashed red lines show the approximations (\ref{Gamma-R-approx})
and (\ref{Gamma-R-lim}) for ${\bar n}\ll n_{\rm crit}$ and ${\bar
n}\gg n_{\rm crit}$, respectively. The dot-dashed black line (almost
indistinguishable from the lowest blue line) shows the approximation
(\ref{Gamma-large-n}). In the approximations we assume large
$\Delta/\textit{g}$, so that $\Gamma_P\approx \Gamma_d$. }
 \label{fig-Gamma-norm}
\end{figure}

From Eq.\ (\ref{Gamma-large-n}) we see that the Purcell rate $\Gamma_R$ decreases
significantly when $\bar{n}$ becomes comparable to $n_{\rm crit}$,
and it continues to decrease with increasing $\bar{n}$, eventually
approaching zero.
   Figure \ref{fig-Gamma-norm} illustrates the dependence of the
Purcell rate $\Gamma_{R}$ (normalized by the no-drive value
$\Gamma_{P}$) on the ratio $\bar n/n_{\rm crit}$ for several values
of the normalized detuning $\Delta/\textit{g}$. We see that the different curves shown in Fig.\ \ref{fig-Gamma} now collapse onto
practically the same line for $|\Delta/\textit{g} | \agt 10$, with a
significant deviation starting only when $|\Delta/\textit{g}|\alt
5$. The approximation (\ref{Gamma-large-n}) is shown by the black dot-dashed line for large $\Delta/\textit{g}$. This approximation also works well (not shown) for the upper curve, $\Delta/\textit{g}=5$, except for the range $\bar n \alt 3$ (correspondingly $\bar n /n_{\rm crit} \alt 0.5$), because of the difference between $\Gamma_d$ and $\Gamma_P$. From Eq.\ (\ref{Gamma-large-n}), the suppression at $\bar n=n_{\rm crit}$ is
$\Gamma_R/\Gamma_P\approx (3+2\sqrt{2})/16= 0.36$ for large
$\Delta/\textit{g}$.

In the strong suppression limit when $\bar{n} \gg n_{\rm crit}$, the
rate (\ref{Gamma-R-n-2}) [or (\ref{Gamma-large-n})] can be approximated as
    \be
    \Gamma_R \approx \Gamma_d \frac{n_{\rm crit}}{4\bar n}
    \left[1+2\sqrt{\frac{n_{\rm crit}}{\bar n}}\right] .
    \label{Gamma-R-lim}\ee
This approximation and the opposite-limit approximation
(\ref{Gamma-R-approx}) are shown by red dashed lines in Fig.\
\ref{fig-Gamma-norm}.

\subsubsection*{Qubit excitation rate}

\begin{figure}[t]
\includegraphics[width=8.5cm]{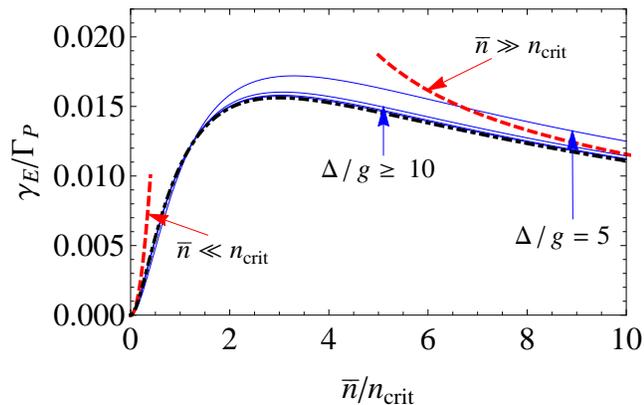}
  \caption{The qubit excitation rate $\gamma_{E}$ [Eq.\
(\ref{gamma-e})] normalized by the no-drive relaxation rate
$\Gamma_{P}$ versus $\bar n/n_{\rm crit}$ for several values of
$\Delta/\textit{g}$: 5, 10, 15, and 20 (blue solid lines). The
curves for $\Delta/\textit{g}\geq 15$ are practically
indistinguishable from each other. The red dashed lines show the
approximations (\ref{gamma-E-approx}) and (\ref{gamma-E-lim}) for
${\bar n}\ll n_{\rm crit}$ and ${\bar n}\gg n_{\rm crit}$,
respectively. The dot-dashed black line (barely distinguishable)
shows the approximation (\ref{gammaE-large-n}). In the
approximations we assume large $\Delta/\textit{g}$.
 }
 \label{fig-gammaE}
\end{figure}

    Similar to the calculation of the qubit energy relaxation rate
(\ref{Gamma-R-n}), we can calculate the qubit excitation rate. Now
the initial state is assumed to be $\overline{|g,n\rangle}$, and the
jump due to the action of the operator $a$ yields the unnormalized
state $a\overline{|gn\rangle}$, which has a non-zero overlap with
the eigenstate $\overline{|e,n-2\rangle}$, corresponding to the
excited state of the qubit. Therefore, the qubit excitation rate is
    \be
 \gamma_{E} (n)= \kappa|\overline{\langle e,n-2|}
 a\overline{|g,n\rangle}|^2,
    \label{gamma}\ee
which can be written as [see Eqs.\
(\ref{en-bar})--(\ref{tan-2theta})]
    \be
 \gamma_{E} (n)=\kappa (\sqrt{n-1}\sin\theta_n\cos\theta_{n-1}
 -\sqrt{n}\sin\theta_{n-1}\cos\theta_n )^2.
    \label{gamma-explicit}\ee
Similar to Eq.\ (\ref{Gamma-R}) this rate should be averaged,
   \be
  \gamma_{E} = \left [ \sum_{n=0}^{\infty} P(n) \, \gamma_{E}(n) \right] / \sum_{n=0}^{\infty} P(n),
    \label{gamma-e}
    \ee
over the probability distribution $P(n)$ for the states $\overline{|g,n\rangle}$, for which we will use the coherent-state approximation (\ref{coh-distr}) with mean photon number $\bar n$ (a better notation used in the next subsection is $\bar n_g$).
 The dependence of $\gamma_E$ on $\bar n$ is shown in Fig.\
\ref{fig-gammaE} by blue solid lines ($\gamma_E$ is normalized by
$\Gamma_P$ and $\bar n$ is normalized by $n_{\rm crit}$).

    Expanding $\gamma_{E} (n)$ in  powers of $g/\Delta$ and carrying out
the summation (\ref{gamma-e}) produces the series
\begin{align}\label{Gaa1}
 \gamma_{E}&=\kappa\bar n^2\lambda^6\big[1-5\lambda^2(2\bar n+3)\notag\\
 &+\lambda^4(69\bar n^2+276\bar n+159)+...\big],
\end{align}
which has the leading-order approximation
    \be
    \gamma_{E} \approx  \frac{\Gamma_d }{16} \left(\frac{\bar{n}}
    {n_{\rm crit}}\right)^2
     \label{gamma-E-approx}
    \ee
at $1\ll \bar{n} \ll n_{\rm crit}$. This dependence is shown by the
left dashed red line in Fig.\ \ref{fig-gammaE}; it works well only
when $\bar n/n_{\rm  crit}\alt 0.1$. [The truncated expansion
(\ref{Gaa1}) works well until $\bar n/n_{\rm  crit}\alt 0.2$.]
 Approximating Eq.\
(\ref{gamma-explicit}) in the opposite limit,  $\bar{n} \gg n_{\rm
crit}$, gives
    \be
    \gamma_{E} \approx \Gamma_d \frac{n_{\rm crit}}{4\bar n}
    \left[1-2 \left(\frac{n_{\rm crit}}{\bar n}\right)^{1/2} +
    3\left(\frac{n_{\rm crit}}{\bar n}\right)^{3/2}  \right] ,
    \label{gamma-E-lim}\ee
which is shown by the right dashed red line in Fig.\
\ref{fig-gammaE}. The approximation for arbitrary $\bar n/n_{\rm
crit}$, which assumes $\bar n\gg 1$ and $n_{\rm crit} \agt 1$
[derived similar to Eq.\ (\ref{Gamma-large-n})] is
    \be
    \gamma_E \approx \frac{\kappa \textit{g}^2}{4\Delta^2}
    \left(\frac{1}{1+\bar n/n_{\rm crit}}
    -\frac{1}{\sqrt{1+\bar n/n_{\rm crit}}}\right)^2 .
    \label{gammaE-large-n}\ee

Note that for $\bar n \alt n_{\rm crit}$ the excitation rate is much
smaller than the relaxation rate, $\gamma_{E}/\Gamma_{R} \alt (\bar
n/4 n_{\rm crit})^2\ll 1$. However, for $\bar n \gg n_{\rm crit}$ the
relaxation and excitation rates become identical in
the leading order, as follows from Eqs.\ (\ref{Gamma-R-lim}) and
(\ref{gamma-E-lim}) (we were not able to reach this regime in
numerical simulations discussed in Sec.\ IV). The dependence
$\gamma_E ({\bar n})$ has a maximum (Fig.\ \ref{fig-gammaE}), which
for $|\Delta/\textit{g}|>3$ occurs at ${\bar n}\approx 3 n_{\rm
crit}$ [this value follows from Eq.\ (\ref{gammaE-large-n})] Even at this maximum the excitation rate is much smaller than the
no-drive relaxation rate, $\gamma_E/\Gamma_P < 0.02$, as seen in
Fig.\ \ref{fig-gammaE} [the maximum value which follows from the
approximation (\ref{gammaE-large-n}) is $\gamma_E=\Gamma_d/64$].

\vspace{0.3cm}

    Our derivation for the relaxation and excitation rates
$\Gamma_R$ and $\gamma_E$ was based on considering only the ``jump''
processes caused by the operator $a$ and finding their contribution
to transitions between the two ladders of eigenstates in Fig.\
\ref{fig-JC}. The remaining (non-unitary) ``no jump'' evolution and
the unitary evolution due to the drive with amplitude $\varepsilon$
also contribute to transitions between the two ladders of
eigenstates. However, since these transitions are non-stochastic,
the energy shift between the two ladders suppresses the transitions
(as expected for a coherent off-resonant process) and prevents the
linear-in-time increase of the ``wrong'' ladder population. This is
why these processes are not expected to contribute directly to the
relaxation and excitation rates. Nevertheless, the evolution due to
the drive and due to the absence of jumps effectively changes the
ladders of eigenstates. Thus, at very large $\bar n$ our results
(\ref{Gamma-R-n-2}) and (\ref{gamma-explicit}) for the relaxation
and excitation rates are expected to become invalid.

\subsection{Purcell rate with drive: Formal approach}

The results of the previous subsection were based on a physical intuition, leading to Eqs.\ (\ref{Gamma-R-n}) and (\ref{gamma}). Here we present a formal derivation of the analytical expression
for the Purcell rate with the microwave drive. In the formal approach we need to assume the slightly nonlinear dispersive regime: $|\Delta/g|\gg 1$ and $\bar
n/n_{\rm crit}\ll 1$. The derivation essentially follows the
formalism developed in Ref. \cite{Boi09}.

 We first transform the master equation \eqref{mast} to the frame
where the JC Hamiltonian is diagonal. While it is simple for
wavefunctions [see Eqs.\ (\ref{en-bar}) and (\ref{gn-bar})], it is
not so simple for the operator form of the Hamiltonian. This can be
done by introducing the transformation $\textbf{D}$ of the form
    \be
\textbf{D}=e^{-\Lambda (N_{e}) I_{-}},
    \ee
where
    \be
I_{\pm}=\sigma_{+}a\pm\sigma_{-}a^{\dag} , \,\,\,
N_{e}=a^{\dag}a+|e\rangle\langle e|,
    \ee
$N_e$ is the number operator for total excitations in the
system and the function $\Lambda(N_{e})$ still has to be determined.
After this transformation the master equation reads
\be\label{mastd}
 \dot \rho^{\textbf{D}}=-i[H_{I}^{\textbf{D}},\rho^{\textbf{D}}]
  +\kappa\mathcal{D}[a^{\textbf{D}}]\rho^{\textbf{D}},
  \ee
  where
  \be \label{tra}
 \rho^{\textbf{D}}=\textbf{D}^{\dag}\rho\textbf{D}, \,\,\,
  a^{\textbf{D}}=\textbf{D}^{\dag}a\textbf{D}, \,\,\,
  H_{I}^{\textbf{D}} = \textbf{D}^{\dag} H_{I}  \textbf{D},
  \qquad
\ee
and $H_I$ is given by Eq.\ (\ref{ham}) (in our notation the
wavefunctions are transformed with $\textbf{D}^\dagger$ rather than
with $\textbf{D}$). The JC part of the transformed Hamiltonian,
$H_{JC}^{\textbf{D}}=(\Delta /2)
\sigma_{z}^{\textbf{D}}+\textit{g}I_{+}^{\textbf{D}}$, can then be
written as \cite{Boi09}
\begin{align}\label{ham0}
  H_{JC}^{\textbf{D}}&=\left[\frac{\Delta}{2}\cos(2\Lambda\sqrt{N_{e}})
 -\textit{g}\sqrt{N_{e}}\sin(2\Lambda\sqrt{N_{e}})\right]\sigma_{z}
\notag\\
  &+\left[\frac{\Delta}{2\sqrt{N_{e}}}\sin(2\Lambda\sqrt{N_{e}})
  +\textit{g}\cos(2\Lambda\sqrt{N_{e}})\right]I_{+} ,
\end{align}
where $\Lambda=\Lambda(N_e)$. Note that here we treated the operator
$N_e$ in the square root as a {\it c}-number. This is possible
because $N_e$ is a positive operator and its square root is defined
via taking the square root of the eigenvalues in the diagonalizing
basis. Also note that $\Lambda$ commutes with $\sqrt{N_e}$, and
that the $\sqrt{N_e}$ in denominator is canceled when the sine function is expanded.

Since we want to diagonalize the JC Hamiltonian, we need to
eliminate the second line of Eq.\ (\ref{ham0}) by zeroing the
coefficient of $I_{+}$, which can be done by choosing
\begin{equation}\label{angle}
  \Lambda=-\frac{1}{2\sqrt{N_{e}}}\arctan\left(\sqrt{4\lambda^2N_{e}}
  \right ) .
\end{equation}
Then the transformed JC Hamiltonian is only the first line of Eq.\
(\ref{ham0}), which using Eq.\ \eqref{angle} can be written as
\begin{align}\label{ham1}
  H_{JC}^{D}&=\frac{\Delta}{2}\sqrt{1+4\lambda^2 N_{e}}~\sigma_{z}
  =\frac{\Delta}{2}\sqrt{1+N_{e}/n_{\rm crit}}~\sigma_{z}.
\end{align}
This is exactly the desired Hamiltonian for the qubit-resonator
system in the eigenbasis -- see Eq.\ (\ref{E-eigen}).

   Next, we find explicit form of the $\textbf{D}$-transformed
annihilation operator $a$, which enters the drive Hamiltonian and
the Lindblad term of the master equation (\ref{mastd}). Calculating
$\textbf{D}^{\dag}a\textbf{D}$ by expanding the exponent in the
definition of $\textbf{D}$, we obtain
\begin{equation}\label{ad-formula}
  a^{\textbf{D}}=a-[a,\Lambda I_{-}]+\frac{1}{2!}[[a,\Lambda I_{-} ],\Lambda I_{-} ]+ ... \,\, .
\end{equation}
Then by expanding $\Lambda(N_{e})$ [using Eq.\ (\ref{angle})] in
powers of $\sqrt{4\lambda^2 N_{e}}$ with the assumption that
$\sqrt{4\lambda^2 N_{e}}<1$ (i.e. $N_e <n_{\rm crit}$)  and
explicitly computing the resulting commutation relations, we obtain
after a lengthy algebra the following expression,
  \be \label{ad}
a^{\textbf{D}}=a^{\textbf{D}}_{1}+a^{\textbf{D}}_{2}+a^{\textbf{D}}_{3},
  \ee
  with
  \begin{eqnarray}
  \label{aD1}
&& a^{\textbf{D}}_{1}=\Big\{1+\frac{\lambda^2}{2}\sigma_{z}
-\frac{\lambda^4}{8}[12(\hat n+1)\sigma_{z}+1]
     \notag\\
&& \hspace{0.7cm} +\lambda^6 \Big[\Big(5\hat n^2+10\hat n
+\frac{73}{16}\Big)\sigma_{z}+\frac{\hat n+1}{4}\Big]+...\Big\}a,
\qquad
    \\
  \label{aD2}
&& a^{\textbf{D}}_{2}=\lambda\Big\{1-\frac{3}{2}\lambda^2(2\hat n+1)
+\lambda^4\Big(11\hat n^2+11\hat n+\frac{31}{8}\Big)
     \notag\\
&& \hspace{0.7cm} -\lambda^6\Big(42\hat n^3+63\hat
n^2+\frac{355}{8}\hat n+\frac{187}{16}\Big)+...\Big\}\sigma_{-},
    \\
  \label{aD3}
&& a^{\textbf{D}}_{3}=\lambda^3\Big\{1-\frac{5\lambda^2}{2}(2\hat
n+3) +\lambda^4\Big(22\hat n^2+66\hat n+\frac{411}{8}\Big)
      \notag\\
&& \hspace{0.7cm} +...\Big\}a^2\sigma_{+},
\end{eqnarray}
where $\hat n=a^{\dag}a$ is the photon number operator. Note that in
the transformed frame the field operator $a^{\textbf{D}}$ acquires a
qubit part ($a^{\textbf{D}}_{2}$ is proportional to $\sigma_{-} $
and $a^{\textbf{D}}_{3}$ is proportional to $a^2\sigma_{+}$).
Equation \eqref{ad} has the following interpretation. Each term
describes an ``annihilation'' process, which reduces the number of
excitations by 1. The first term $a^{\textbf{D}}_{1}$
(proportional to $a$) describes annihilation of a photon in the
eigenbasis, which is modified due to the presence of the qubit. The
second term $a^{\textbf{D}}_{2}$, which describes qubit relaxation,
also reduces the number of excitations by 1. Thus the photon
annihilation process is partly converted into qubit relaxation: this second term eventually leads to the Purcell relaxation. The last
term $a^{\textbf{D}}_{3}$, which describes annihilation of two
photons and qubit excitation, also reduces the number of excitations
by 1. This process leads to qubit excitation as a result of
leakage of photons through the resonator.
  There are no more groups of terms because qubit cannot absorb or
emit more than one excitation, and thus there are no more processes
decreasing the total number of excitations by 1.

It is interesting to relate the terms in Eq.\ (\ref{ad}) with the
matrix elements of $a$ in the eigenbasis, considered in the previous
subsection. Since $a^{\textbf{D}}$ is essentially the operator $a$
in the eigenbasis, we would expect that $a^{\textbf{D}}$ sandwiched
between two bare states should be equal to the operator $a$
sandwiched between the corresponding eigenstates so that
\begin{eqnarray}
     \label{ev1}
&&   \overline{\langle e,n-1|}a\overline{|e,n\rangle}=
  \langle e,n-1|a^{\textbf{D}}_{1}|e,n\rangle,
  \\
  \label{ev2}
&&  \overline{\langle g,n|}a\overline{|e,n\rangle}=
  \langle g,n|a^{\textbf{D}}_{2}|e,n\rangle,
    \\
  \label{ev3}
&&  \overline{\langle e,n-2|}a\overline{|g,n\rangle}=
  \langle e,n-2|a^{\textbf{D}}_{3}|g,n\rangle.
\end{eqnarray}
  We have checked these relations explicitly using the truncated
expansions (\ref{aD1})--(\ref{aD3}) and corresponding expansions of
the eigenstates (\ref{en-bar})--(\ref{tan-2theta}). These relations
give us an insight why the formal-approach results which we will
obtain later are essentially equivalent to the results of the simple
approach.

The next step in the derivation is to use the polaron-type
transformation described below.

\subsubsection*{Polaron-type transformation}

   We expect that in the eigenbasis the quasi-stationary state is
 close to a coherent state within the subspace of ``excited''
 eigenstates (right ladder of red dashed lines in Fig.\ \ref{fig-JC})
 and also close to a (possibly different) coherent state within the
 subspace of ``ground state'' eigenstates (left ladder of dashed
 lines). Therefore, it is natural to apply displacement
 transformations within these (eigen) ladders, which transform the
 coherent states to the lowest levels in the ladders. In this way
 the case with the microwave drive should, to a considerable
 extent, be reduced to the case without the microwave drive,
 thus simplifying the analysis.

    Formally, we apply the polaron-type \cite{Gam08} transformation
$\textbf{P}$ to the master equation \eqref{mastd}, so that the
density operator transforms as
$\rho^{\textbf{DP}}=\textbf{P}^{\dag}\rho^{\textbf{D}}\textbf{P}$
with
    \begin{equation}\label{P}
\textbf{P}=|e\rangle\langle e|D(\alpha_{e}) +|g\rangle\langle
g|D(\alpha_{g}),
\end{equation}
where $D(\alpha)$ is the usual displacement operator and
$\alpha_{e(g)}$ are resonator field amplitudes corresponding to the
qubit state $|e\rangle$ or $|g\rangle$. Note that Eq.\ (\ref{P})
formally uses the bare states, but the conversion between bare states and eigenstates is already performed by the transformation $\textbf{D}$. Also note that we
apply $\textbf{P}^{\dag}$ (not $\textbf{P}$) to the wavefunctions,
so this is a displacement by $-\alpha_{e(g)}$.  So far the
amplitudes $\alpha_{e}(t)$ and $\alpha_{g}(t)$ are arbitrary and in
general time-dependent. The master equation after the polaron-type
transformation becomes
\begin{align}\label{ma}
\hspace{-0.05cm}  \dot \rho^{\textbf{DP}}=&-i[H_{JC}^{\textbf{DP}}+
  \varepsilon(a^{\textbf{DP}}+a^{\dag \textbf{DP}}), \rho^{\textbf{DP}}]
  +\kappa\mathcal{D}[a^{\textbf{ DP}}]\rho^{\textbf{DP}}\notag\\
  &+i[\text{Im}(T_{\alpha}\dot{T}_{\alpha}^{*}),\rho^{\textbf{DP}}]
  +[a\dot{T}_{\alpha}^{*}-a^{\dag}\dot{T}_{\alpha},\rho^{\textbf{DP}}],
  \end{align}
where $T_{\alpha}=\alpha_{e}|e\rangle\langle
e|+\alpha_{g}|g\rangle\langle g|$ and the second line in this
equation is due to time-dependence of $\alpha_{e}(t)$ and
$\alpha_{g}(t)$.
  The explicit form of the master equation (\ref{ma}) is very
lengthy and we do not present it here. Its perturbative form can be
obtained by first expanding the Hamiltonian $H_{JC}^{\textbf{D}}$ in
powers of $N_{e}/n_{\rm crit}$ [see Eq.\ (\ref{ham1})] and then
applying the $\textbf{P}$-transformation to the field and qubit
operators,
    \begin{eqnarray}\label{a}
&&   \textbf{P}^{\dag}a {\textbf{P}}=a+T_{\alpha},
  \\
&&  \sigma_{z}^{\textbf{P}}=\sigma_{z}, \,\,\,
\textbf{P}^{\dag}\sigma_{-}
  {\textbf{P}}=D(\alpha_{g})^{\dag}D(\alpha_{e})\sigma_{-} ,
  \end{eqnarray}
while the transformation of higher-order terms for field operators
can be obtained using sequential application of \eqref{a}, for
example,
\begin{eqnarray}
&&  \textbf{P}^{\dag}\hat n{\textbf{P}}=\hat
n+aT_{\alpha}^{*}+a^{\dag}T_{\alpha}+|T_{\alpha}|^2,
  \\
&&  \textbf{P}^{\dag} \hat n^2
{\textbf{P}}=[(1+2|T_{\alpha}|^2)T_{\alpha}^{*}a+a^2T_{\alpha}^{*2}
+2\hat na T_{\alpha}^{*}+\text{H.c.}]
 \notag\\
  && \hspace{1.5cm} +|T_{\alpha}|^2(4\hat n+1)+\hat n^2+|T_{\alpha}|^4.
\end{eqnarray}

Now we want to find ``good'' values of $\alpha_e$ and $\alpha_g$,
which correspond to the quasi-stationary state. This can be done
using the following trick. Let us impose the condition that the
total coefficient for the operator $a^{\dag}$ in the transformed
master equation \eqref{ma} is zero (then the coefficient for $a$
also vanishes automatically). This would correspond to the situation
without drive (in the effective frame), and then because of the
relaxation due to $\kappa$, the lowest state within each ladder will
be eventually reached, independent of the initial state. [This will
not be exact because of non-zero terms $(a^\dagger)^2$,
$(a^\dagger)^3$, etc., but this is good as an approximation.]
Imposing this condition and keeping terms up to $\lambda^4$, we
obtain equations
\begin{align}\label{al}
 \dot\alpha_{j}(t)\approx &-\frac{\kappa}{2}
 \left[1\pm\lambda^2(1-6\lambda^2 \bar n_{j})\right]\alpha_{j}
  \notag\\
  &+i\chi\{\lambda^2\mp  [1-2\lambda^2(\bar n_{j}+1)]\}\alpha_{j}
  \notag\\
  &-i\varepsilon\left\{1-\frac{\lambda^4}{8}\pm \frac{\lambda^2}{2}
  \left[1-3\lambda^2(2\bar n_{j}+1)\right]\right\},
\end{align}
with the top sign for $j=e$ and the bottom sign for $j=g$; here
$\bar n_{j}=|\alpha_{j}|^2$ is the corresponding mean photon number
and $\chi=g^2/\Delta$ is the resonator frequency shift in the strong
dispersive regime ($\lambda\ll 1$, $\bar n_{j} \ll n_{\rm crit}$).
In Eq.\ \eqref{al} the term proportional to $\kappa$ is a
contribution from the Lindblad master equation, the term
proportional to $\chi$ is a contribution from JC Hamiltonian, and
the last term is due to the microwave drive.
  It is easy to see that Eq.\ \eqref{al} is essentially the equation
for classical field amplitudes $\alpha_{e(g)}(t)$, as expected. The
stationary solution of this equation, $ \dot\alpha_{j}(t)=0$, gives
the steady-state values $\alpha_e$ and $\alpha_g$,
 which are then substituted into the master equation (\ref{ma}).

    With these ``good'' values of $\alpha_e$ and $\alpha_g$,
we expect a significant population of only two states in the
$\textbf{D}\textbf{P}$-transformed frame: $|e,0\rangle$ and
$|g,0\rangle$. We also expect that these populations,
$\rho_{e0,e0}^{\textbf{DP}}$ and $\rho_{g0,g0}^{\textbf{DP}}$, are
close to the total occupation of the right and left ladders of
eigenstates  in Fig.\ \ref{fig-JC}. Therefore the transition rates
between the states $|e,0\rangle$ and $|g,0\rangle$ in the
$\textbf{D}\textbf{P}$-transformed frame should give the relaxation
and excitation rates for the qubit; these rates can now be found
from Eq.\ (\ref{ma}). The expansion form of the equation for
$\dot{\rho}_{e0,e0}^{\textbf{DP}}$ is very lengthy and we do not
show it here, but if we keep only the terms with
$\rho_{e0,e0}^{\textbf{DP}}$ and $\rho_{g0,g0}^{\textbf{DP}}$, we then obtain
\begin{equation}\label{eq}
  \dot \rho_{e0,e0}^{\textbf{DP}}\approx
  -\Gamma_{R}\rho_{e0,e0}^{\textbf{DP}}
  +\gamma_{E}\rho_{g0,g0}^{\textbf{DP}},
\end{equation}
where (to order $\lambda^8$) we find
\begin{align}\label{gt}
 \Gamma_{R}&=\kappa \lambda^2\big[1-3\lambda^2(2\bar n_{e}+1)+\lambda^4(31 \bar n^2_{e}+62\bar n_{e}+10)\notag\\
                &-\lambda^6(150\bar n_{e}^3+675\bar n_{e}^2+520 \bar n_{e}+35)\big],
\end{align}
and (to order $\lambda^{10}$)
\begin{align}\label{Gab1}
 \gamma_{E}=\kappa \bar n_{g}^2\lambda^6\big[1-5\lambda^2(2\bar n_{g}+3)
+\lambda^4(69\bar n_{g}^2+276\bar n_{g}+159)\big].
\end{align}
These results for the relaxation and excitation rates $\Gamma_R$ and
$\gamma_E$ are in exact agreement with Eqs.\ \eqref{Gamma-R-exp} and
\eqref{Gaa1}, obtained using the simple intuitive approach. Note however that
since our formal derivation was based on the expansion in $\lambda$,
it can be used only for relatively small values of the nonlinearity
parameter $\bar n_{e(g)}/n_{\rm crit}$ (see, e.g., the green dashed line
in Fig.\ \ref{fig-Gamma}), in contrast to the simple approach.

 Thus the results (\ref{gt}) and (\ref{Gab1}) of the formal approach
confirm that the Purcell relaxation rate decreases with increasing strength
of the microwave drive, and there exists a relatively weak qubit
excitation due to resonator damping. The results of the formal
approach are essentially an extension of the previous work
\cite{Boi09}, where only the leading order was considered, and hence
the photon-number-dependent Purcell rate and the excitation rate
were not obtained.

\subsection{Physical interpretation}

    We do not have a compete physical interpretation of the Purcell
relaxation suppression due to additional photons induced by the
microwave drive, so we may say that this is just the mathematical property of the
matrix element $\overline{\langle g,n|}a\overline{|e,n\rangle}$ [see
Eq.\ (\ref{Gamma-R-n})], which decreases with $n$. However, we have
a crude physical interpretation. The idea is that the additional
drive changes the effective qubit frequency due to the ac (dynamic)
Stark shift, which increases the detuning $|\Delta|$. Then if we use
the formula for the single-excitation Purcell rate [Eqs.\
(\ref{gcd}) or (\ref{ga})], we find the suppression of the rate with
increasing mean photon number $\bar n$.

\begin{figure}[t]
\includegraphics[width=8cm]{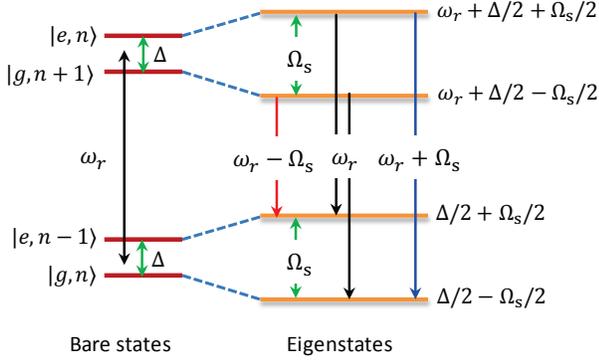}
  \caption{Four neighboring levels of the Jaynes-Cummings ladder
of Fig.\ \ref{fig-JC}, redrawn in a slightly different way. The
shown energy is relative to the energy of the state $|g,n\rangle$.
Assuming $n\approx {\bar n}\gg 1$, the level splitting in both pairs
of eigenstates is approximately $|\Omega_S| =\sqrt{\Delta^2
+4\textit{g}^2\bar n}$. The level splitting increases with $\bar n$.
There are four possible transitions marked by down arrows; two of
them have frequencies almost coinciding with $\omega_{\rm r}$
(neglecting the dispersive shift), and two of them are at
frequencies $\omega_{\rm r}\pm \Omega_S$. These transitions can be
thought of as the Mollow triplet \cite{Mol69} at non-zero detuning.
}
 \label{fig-ds}
\end{figure}

    More quantitatively, the effective detuning with ${\bar n}\gg 1$
photons is the level splitting between the eigenstates of the JC
ladder (see Figs.\ \ref{fig-JC} and \ref{fig-ds}),
    \be
    \Omega_S =\sqrt{\Delta^2+4{\bar n}\textit{g}^2}\, {\rm
sgn}(\Delta)=\Delta \sqrt{1+{\bar n}/n_{\rm crit}} .
    \label{Omega-S}\ee
(Note that the resonator frequency changes by less than
$\textit{g}^2/\Delta=\Delta/4n_{\rm crit}$, so the change of
effective detuning is mostly due to the qubit frequency change.) In
the regime ${\bar n}\ll n_{\rm crit}$, this gives the effective
detuning $\Delta (1+{\bar n}/2n_{\rm crit})$. Therefore, if we use
the dispersive single-excitation formula $\Gamma_d=\kappa
\textit{g}^2/\Delta^2$ [Eq.\ (\ref{gcd})] with increased $|\Delta|$,
we would expect the Purcell rate suppression as $\Gamma_R \approx
\Gamma_d (1-{\bar n}/n_{\rm crit})$, which is different from the
actual result [Eq.\ (\ref{Gamma-R-approx})] by the missing factor
$3/2$. The additional suppression could in principle be explained by
the change of effective $\kappa$ (due to decreasing overlap);
however, we did not find a reasonable quantitative explanation of
this kind.
    Besides the difficulty with quantitative explanation of the
suppression, a natural question is why we can still use the no-drive
formula (\ref{gcd}), essentially mixing the linear and nonlinear
approaches to the dynamics. Thus we cannot call the physical
interpretation based on the ac Stark shift a perfect interpretation.

    Note that it is also possible to discuss the reduction of the
qubit relaxation rate and appearance of the nonzero excitation rate in
terms of the Mollow triplet physics \cite{Mol69} at non-zero
detuning  $\Delta$ (Fig.\ \ref{fig-ds}). In our case the triplet is
transformed into a quadruplet (neglecting the ``fine structure'' due to $n$-dependence of the transition frequencies).
The two central ``peaks'' correspond to the qubit-state-dependent resonator
frequency $\omega_{\rm r} \pm \textit{g}^2/\Omega_S$; the difference between them is used for the qubit readout \cite{Blais-04,Joh12,Ris12,X-mon13} (the quadruplet becomes a triplet if the dispersive shift $\pm \textit{g}^2/\Omega_S$ is neglected, as in Fig.\ \ref{fig-ds}). The two side peaks correspond
to the qubit relaxation and excitation (Fig.\ \ref{fig-ds}). In the
case of relaxation the qubit emits the photon with frequency
approximately $\omega_{\rm r}+ \Omega_S$. In the case of excitation
the qubit absorbs this energy, which is taken from two photons in
the resonator, so that the photon emitted into the transmission line
has frequency $\omega_{\rm r}- \Omega_S$. With increasing drive
(increasing $\bar n$) the side peaks move further away from the
central peaks (ac Stark shift). As discussed above, this leads to
the suppression of the relaxation rate $\Gamma_R$ (at least
qualitatively). The excitation rate $\gamma_E$ first increases with
increasing $\bar n$, as expected for a two-photon process, but
eventually the suppression due to increasing detuning becomes the
dominating effect, and $\gamma_E$ starts to decrease.

\section{Numerical results}

Besides developing the analytical approaches discussed in the
previous section, we have also solved the full master equation
(\ref{mast}) for the qubit-resonator system numerically and thus
computed the qubit relaxation and excitation rates. In this section
we present the numerical results and compare them with analytical
results.

\subsection{Qubit relaxation rate}

Since we do not consider intrinsic qubit relaxation as well as
dressed dephasing (a process which converts pure depahsing into photon-number-dependent qubit relaxation), the decay rate obtained from the numerical
solution is only due to the Purcell effect. To calculate the
relaxation rate we start with an initial state that is the
coherent-state superposition of the eigenstates corresponding to the
excited state of the qubit \cite{Set13} (right ladder of red dashed
lines in Fig.\ \ref{fig-JC}),
    \begin{equation}\label{in}
  |\psi_{\text{in}}\rangle=e^{-|\alpha_{\text{in}}|^2/2}
  \sum_{n=0}^{\infty}\frac{\alpha_{\text{in}}^n}{\sqrt{n!}} \,
  \overline{|e,n\rangle},
\end{equation}
where $\alpha_{\rm in}$ is the initial amplitude of the resonator
field. For $\alpha_{\rm in}$ we use the value
 $\alpha_{\rm in}=-i\varepsilon/[i(\textit{g}^2/\Delta +\omega_{\rm r}-\omega_{\rm d})
+\kappa/2]$, obtained classically by assuming that the resonator
frequency is increased by $\textit{g}^2/\Delta$ due to the qubit in
the excited state (recall that the driving frequency $\omega_{\rm
d}$ is exactly $\omega_{\rm r}$).
 The corresponding mean photon number is
${\bar n}_{\rm in} = |\alpha_{\text{in}}|^2=4|\varepsilon|^2/[4
   (\textit{g}^2/\Delta +\omega_{\rm r}-\omega_{\rm d})^2+\kappa^2]$.
However, this initial value of $\alpha_{\rm in}$ is good only in the
linear regime when $\bar n \ll n_{\rm crit}$. In the non-linear
regime the effective resonator frequency becomes (see Fig.\
\ref{fig-JC}) $\omega_{\rm r}+ \Delta (\sqrt{1+{\bar
n}/n_{crit}}-\sqrt{1+({\bar n}-1)/n_{\rm crit}}) \approx \omega_{\rm
r}+\textit{g}^2/\Omega_S$, where the level splitting $\Omega_S$ is
given by Eq.\ (\ref{Omega-S}). Therefore the quasi-steady state is
expected to have $\alpha_e \approx
-i\varepsilon/[i(\textit{g}^2/\Omega_S +\omega_{\rm r}-\omega_{\rm d}) +\kappa/2]$ and the
corresponding mean photon number is expected to be
 \begin{equation}\label{mn}
   \bar n=\frac{|\varepsilon|^2}
   {(g^2/\sqrt{\Delta^2+4g^2\bar n}
   +\omega_{\rm r}-\omega_{\rm d})^2+(\kappa/2)^2}.
\end{equation}
This equation allows easy calculation of $\varepsilon$ for a desired
$\bar n$. Note that for the qubit in the ground state, the frequency shift  $\textit{g}^2/\Omega_S$ is replaced with $-\textit{g}^2/\Omega_S$. However, since we use $\omega_{\rm d}=\omega_{\rm r}$, the mean photon number $\bar n$ given by Eq.\ (\ref{mn}) does not change. Therefore,  $\bar n$
is not affected by the Purcell relaxation.
  Also note that Eq.\ (\ref{mn}) does not show bistability when $\omega_{\rm d}=\omega_{\rm r}$, though the bistability
occurs for a range of detuning $\omega_{\rm d}-\omega_{\rm r}\neq 0$
if $\kappa < 4\textit{g}^4 {\bar n}/\Omega_S^3$.
   In simulations we
first find $\varepsilon$ for a desired $\bar n$ analytically, but
then calculate the actual $\bar n$ numerically. We have checked that
Eq.\ (\ref{mn}) works quite well for the parameters we used, but
still not perfectly.

    We use the bare basis to compute the evolution using the master
equation (\ref{mast}), but then convert the results into the
eigenbasis; in particular, we monitor the population of the exited
qubit state in the eigenbasis,
    \be
    \overline{\rho}_{ee} (t)= \sum\nolimits_n {\rm Tr}
  [\rho (t) \, \overline{|e,n\rangle}\,\, \overline{\langle e,n|}\, ] .
    \ee
The use of the initial condition (\ref{in}) allows us to mostly
avoid initial oscillations of  $\overline{\rho}_{ee} (t)$ (decaying
on the timescale $\sim \kappa^{-1}$), so that the decay of
$\overline{\rho}_{ee}$ is smooth in time, and therefore the Purcell
rate $\Gamma_R$ can be relatively easily defined numerically as the
slope of the dependence $-\ln [\overline{\rho}_{ee}(t)]$, which is
close to a straight line for a sufficiently long duration. Note that
because of a non-zero excitation rate, the dependence
$\overline{\rho}_{ee}(t)$ eventually saturates at a non-zero value;
therefore it is necessary to restrict the time duration, making sure
that $\overline{\rho}_{ee}$ remains much larger than the saturation
value.

\begin{figure}[t]
\includegraphics[width=7cm]{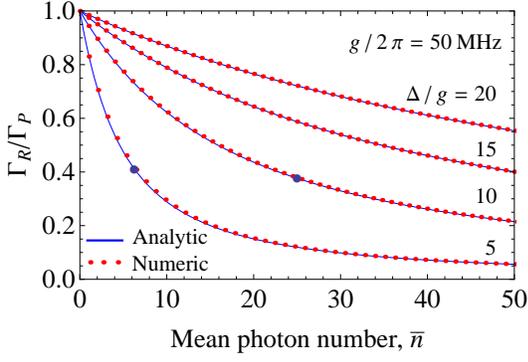}
\caption{Normalized Purcell rate $\Gamma_{R}/\Gamma_{P}$ vs the mean
photon number for $\kappa=\textit{g}=2\pi \times
 50$ MHz and several values of detuning: $\Delta/g=5$, 10, 15, and 20.
The red dots show the results obtained numerically, blue solid lines
are calculated using Eq.\ (\ref{Gamma-R}). The large blue dots
indicate the critical photon number $n_{\rm crit}$. }
   \label{fig-Gamma-num}
\end{figure}

   Red dots in Fig.\ \ref{fig-Gamma-num} show the numerically calculated Purcell
rate $\Gamma_R$ [normalized by the no-drive value $\Gamma_P$ given
by Eq.\ (\ref{ga})] as a function of the steady-state mean photon
number $\bar n$ in the resonator, for $\kappa = \textit{g}$ and
several values of $\Delta/\textit{g}$ (the calculations have been
done for $\textit{g}/2\pi =50$ MHz, but the results do not depend on
$\textit{g}$ because of the linear overall scaling). The solid lines
show the approximate analytical result, obtained in the simple
approach, Eq.\ (\ref{Gamma-R}). We see a very good agreement between
the analytics and numerics. It is important that the Purcell rate
continues to decrease when $\bar n$ is significantly larger than
$n_{\rm crit}$, and agreement with analytics is still very good in
this regime. The numerical results confirm that the Purcell rate
suppression can be more than an order of magnitude.

\begin{figure}[t]
\includegraphics[width=7cm]{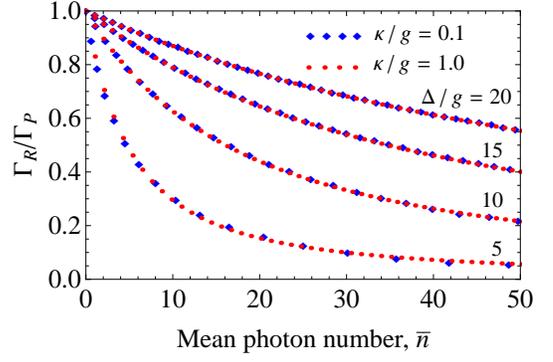}
\caption{Numerical results for the normalized Purcell rate
$\Gamma_{R}/\Gamma_{P}$ as a function of $\bar n$ for
$\kappa/\textit{g}=0.1$ (blue diamonds) and $\kappa/\textit{g}=1$
(red dots); the detuning parameter is $\Delta/g=5$, 10, 15, and 20
(as in Fig.\ \ref{fig-Gamma-num}).}
   \label{fig-Gamma-num2}
\end{figure}

The analytics [Eqs.\ (\ref{Gamma-R-n})--(\ref{Gamma-R})] predict
that the Purcell rate $\Gamma_R$ scales linearly with $\kappa$
(keeping the same $\bar n$). In Fig.\ \ref{fig-Gamma-num2} we check
this scaling numerically by comparing the results for
$\kappa/\textit{g}=1$ (red dots, same as in Fig.\
\ref{fig-Gamma-num}) and for $\kappa/\textit{g}=0.1$ (blue diamonds).
(Note that $\kappa/\textit{g}=0.1$ and even lower values are typical
in circuit QED experiments). We see that the numerical results
confirm, at least in this regime, the simple scaling of $\Gamma_R$
with $\kappa$ (so that $\Gamma_R/\Gamma_P$ does not depend on
$\kappa$) for the same value of $\bar n$. Note however that
decreasing $\kappa$ assumes decreasing drive amplitude $\varepsilon$
to keep $\bar n$ fixed.

\subsection{Qubit excitation rate}

We have also calculated numerically the qubit excitation rate due to
the resonator decay. For that we start with the initial state, which
is the superposition of the eigenstates corresponding to the ground
state of the qubit:
 \begin{equation}\label{ing}
  |\psi_{\text{in}}\rangle=e^{-|\alpha_{\rm in}|^2/2}
  \sum_{n=0}^{\infty}\frac{\alpha_{\text{in}}^n}{\sqrt{n!}} \,
  \overline{|g,n\rangle},
\end{equation}
where $\alpha_{\rm in}$ now corresponds to the ground state of the
qubit, $\alpha_{\rm in}=-i\varepsilon/[i(-\textit{g}^2/\Delta
 +\omega_{\rm r}-\omega_{\rm d})+
\kappa/2]$ or in a better approximation $\alpha_g
=-i\varepsilon/[i(-\textit{g}^2/\Omega_S  +\omega_{\rm r}-\omega_{\rm d})+ \kappa/2]$. Since we use  $\omega_{\rm d}=\omega_{\rm r}$, only the phase
of $\alpha_{\rm in}$ is different from what was used in Eq.\
(\ref{in}), while $\bar n_{\rm in}$ is still the same. Note that the
initial value $\alpha_{\rm in}$ is not very important because the
system converges to the quasi-steady-state value of $\alpha$ within
few resonator lifetimes $\kappa^{-1}$. Similar to what was discussed
in the previous subsection, we calculate the quasi-steady-state
value of $\bar n$ numerically.

\begin{figure}[t]
\includegraphics[width=7.5cm]{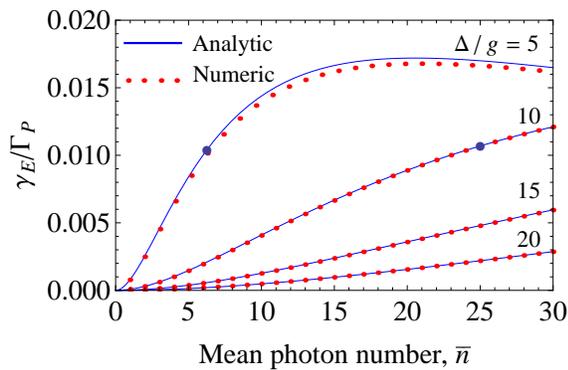}
  \caption{The qubit excitation rate $\gamma_{E}$ (normalized by the
  no-drive relaxation rate $\Gamma_P$) as a function of $\bar n$ for
$\kappa=\textit{g}$ (we used $\textit{g}/2\pi=50$ {MHz}) and several
values of detuning, $\Delta/g$=5, 10, 15, and 20. The red dots show
the numerical results, the blue solid lines are obtained analytically
using Eq.\ (\ref{gamma-e}). The large blue dots indicate $n_{\rm
crit}$.  }
 \label{fig-gammaE-num} \end{figure}

Following the same procedure as for computing the qubit relaxation
rate, we calculate the qubit excitation rate $\gamma_E$. The only
difference is that now we monitor the decrease of
$\overline{\rho}_{gg}(t)=1-\overline{\rho}_{ee}(t)$, and the qubit
excitation rate $\gamma_{ E}$ is defined numerically as the negative
slope of $\ln [\overline{\rho}_{gg}(t)]$ during a time range when
this dependence is linear (this time range should be much shorter
than $\Gamma_R^{-1}$).

    Figure \ref{fig-gammaE-num} shows the excitation rate
$\gamma_{E}$ normalized by the no-drive Purcell rate $\Gamma_{P}$
[Eq.\ (\ref{ga})], as a function of the mean photon number $\bar n$.
The excitation rate is quite weak and first increases with the
strength of the microwave field. The rate also depends on the
detuning: the smaller the detuning, the larger the excitation rate
becomes. The rate calculated numerically agrees well with the
analytical result given by Eqs.\ \eqref{gamma-e}. As discussed in
Sec.\ IIIA, the dependence $\gamma_E(\bar n)$ should reach maximum
at $\bar n \approx 3 n_{\rm crit}$ (when $\Delta/\textit{g}>3$).
Figure \ref{fig-gammaE-num} confirms this behavior for
$\Delta/\textit{g}=5$. Even at the maximum, the value of $\gamma_E$
remains much smaller than $\Gamma_P$ (over 50 times), and at this
point it is $\sim 8$ times smaller than $\Gamma_R$.

\section{Conclusion}

We have analyzed the Purcell relaxation rate of a superconducting
qubit coupled to a leaking microwave resonator, which is pumped
on-resonance by an external microwave drive. The main result is that
the Purcell rate $\Gamma_R$ is suppressed due to the presence of
photons in the resonator, with the strong suppression obtained in
the nonlinear regime.
    The presence of photons in the resonator also leads to qubit
excitation, but the excitation rate $\gamma_E$ is always much
smaller than the no-drive Purcell relaxation rate $\Gamma_P$.
    We have derived approximate analytical formulas for the
relaxation and excitation rates [e.g., Eqs.\ (\ref{Gamma-R-n-2}), (\ref{Gamma-large-n}), (\ref{gamma-explicit}), and (\ref{gammaE-large-n})], which agree well with the numerical results.

In this work we assumed a time-independent drive amplitude $\varepsilon$. It is rather simple to introduce a time-dependent drive $\varepsilon (t)$ to describe experiments with short measurement time, as long as the dynamics is sufficiently adiabatic (which is almost always the case in experiments -- see Ref.\ \cite{Set13}). Using the simple intuitive approach (Sec.\ IIIA), we first solve the classical equations for the field with account of nonlinearity to find $\bar n(t)$ and then obtain the corresponding time-dependent Purcell rate $\Gamma_R (t)$ and the excitation rate $\gamma_E(t)$. In the formal perturbative approach (Sec.\ IIIB) the evolution of the field can be taken into account automatically via Eq.\ (\ref{al}).

To experimentally observe the Purcell rate suppression with
increasing microwave drive predicted in this work, it is important
to make sure that other mechanisms do not increase the qubit
relaxation rate faster than the obtained suppression. One such
mechanism for superconducting qubits is the dressed dephasing
\cite{Boi-08,Boi09,Sli12}, which increases the relaxation rate as
$\Gamma_{\downarrow , {\rm dd}}\simeq \gamma_{\varphi}\bar n/2n_{\rm
crit}$, where $\gamma_{\varphi}$ is the pure
dephasing rate (this formula assumes a similar noise spectrum at the
qubit frequency $\omega_{\rm q}$ and at low frequency). This
increase is weaker than the first-order Purcell rate suppression
$-3\bar n\Gamma_{P}/2n_{\rm crit}$ [see Eq.\ \eqref{Gamma-R-exp}] if
the no-drive Purcell rate $\Gamma_P\approx \Gamma_d=\kappa
\textit{g}^2/\Delta^2$ exceeds the pure dephasing, $\Gamma_P
>\gamma_\varphi /3$. Therefore, to observe the Purcell rate
suppression experimentally, it may be necessary to deliberately
increase the Purcell rate by decreasing the detuning $\Delta$ and/or
increasing the resonator damping rate $\kappa$ and coupling $\textit{g}$.
Note an indication of possible Purcell rate suppression in a recent experiment -- see Fig.\ S7 in Ref.\ \cite{Ris12}.

In this paper we treated the qubit as a two-level system. In
reality, most of present-day superconducting qubits are essentially
only slightly nonlinear oscillators, with almost equidistant energy
spectrum. The anharmonicity
$\mathcal{A}=2E_{|e\rangle}-E_{|g\rangle}-E_{|f\rangle}$ (with
$|f\rangle$ being the next excited level) is typically much smaller
than the qubit frequency ($\mathcal{A}/\omega_{\rm q}\simeq
0.03-0.05$) and only a few times larger than the coupling
$|\textit{g}|$. The presence of the level $|f\rangle$ does not
affect the no-drive Purcell rate $\Gamma_P$; however, it affects the
qubit relaxation and excitation rates when the microwave drive is
applied. The analysis of this effect is beyond the scope of this
paper, but we have done preliminary calculations based on the
natural extension of Eqs.\ (\ref{Gamma-R-n}) and (\ref{gamma}),
assuming that they are still applicable. These calculations show
that the Purcell rate is still suppressed with increasing photon
number $\bar n$ in the typical regime ($\mathcal{A}>0$, $\omega_{\rm
q} <\omega_{\rm r}$), though the suppression is weaker than in the
two-level model. We have also found that the suppression can still
be crudely (not quantitatively) described as being due to the ac
Stark shift of the qubit frequency.

In the slightly nonlinear regime the repulsion of the level
$|e,n\rangle$ from the level $|f,n-1\rangle$ (which is added to the
repulsion between the levels  $|e,n\rangle$ and  $|g,n+1\rangle$)
leads to the effective detuning $\Delta_{\rm eff}=\Delta + 2(n+1)
\textit{g}^2/\Delta - n
(\sqrt{2}\,\textit{g})^2/(\Delta-\mathcal{A})$, where
$\sqrt{2}\,\textit{g}$ is the approximate coupling constant for the
transitions between $|e\rangle$ and $|f\rangle$. This can be
rewritten as
    \be
    \Delta_{\rm eff}=\Delta \left( 1+ \frac{n}{2\tilde{n}_{\rm crit}}
    \right),
    \,\,\, \tilde{n}_{\rm crit} =\frac{\Delta^2}{4 \textit{g}^2} \,
    \frac{\mathcal{A}-\Delta}{\mathcal{A}},
    \label{n-crit-tilde}\ee
where $\tilde{n}_{\rm crit}$ is the appropriately redefined critical
photon number (\ref{n-crit}), which takes into account the third
level, and we assumed $1\ll n\ll n_{\rm crit} \times \min [1,
(\Delta-\mathcal{A})^2/2\Delta^2]$ (so that both level repulsions
are in almost linear regimes). Note that the effective
qubit-state-dependent change of the resonator frequency (which is
used for the qubit measurement) is governed by the similar factor,
$\omega_{{\rm r},|e\rangle}^{\rm eff} -\omega_{{\rm
r},|g\rangle}^{\rm eff} = d\Delta_{\rm
eff}/dn=\Delta/(2\tilde{n}_{\rm crit})$ for the same range of $n$.
 Now crudely estimating the Purcell rate as $\Gamma_R \simeq
\kappa (\textit{g}/\Delta_{\rm eff})^2$, we obtain the crude
estimate of the Purcell rate suppression as $\Gamma_R/\Gamma_P\simeq
1 - n/\tilde{n}_{\rm crit}$ when $1\ll n\ll n_{\rm crit} \times \min
[1, (\Delta-\mathcal{A})^2/2\Delta^2]$. This has the same form as
Eq.\ (\ref{Gamma-R-approx}), with $n_{\rm crit}$ replaced by
$\tilde{n}_{\rm crit}$ and with absent factor $3/2$, which cannot be
obtained in this crude derivation (preliminary numerical results
indicate that this factor, describing the difference between the
numerical suppression and the Stark-shift model becomes closer to
1).
 Note that for positive $\mathcal{A}$ and negative $\Delta$
 (i.e.\ $\omega_{\rm q}< \omega_{\rm r}$,
which is the more typical situation for transmon qubits) the
effective critical photon number (\ref{n-crit-tilde}) is increased,
$\tilde{n}_{\rm crit}
> n_{\rm crit}$ (strongly increased if $-\Delta
\gg \mathcal{A}$), and therefore the Purcell rate suppression is
weaker than in the two-level model. (For large ratios
$-\Delta/\mathcal{A}$ preliminary numerical results indicate that
the suppression is not as weak as follows from strongly increased
$\tilde{n}_{\rm crit}$.) A more complete analysis of the effect of
the third and higher levels on the Purcell rate (including the
numerical simulations as in Sec.\ IV) is the subject of a further
research.

\begin{acknowledgements}
We thank Howard Carmichael, Klaus M{\o}lmer, Daniel Sank, and John
Martinis for useful discussions. We also thank Justin Dressel for a critical  reading of the manuscript. This research was funded by the
Office of the Director of National Intelligence (ODNI), Intelligence
Advanced Research Projects Activity (IARPA), through the Army
Research Office Grants No. W911NF-10-1-0334 and No. W911NF-10-1-0324. All
statements of fact, opinion or conclusions contained herein are
those of the authors and should not be construed as representing the
official views or policies of IARPA, the ODNI, or the U.S.
Government. We also acknowledge support from the ARO MURI Grant No.
W911NF-11-1-0268.
\end{acknowledgements}


\begin{thebibliography}{99}

\bibitem{Pur46} E. M. Purcell, Phys. Rev. \textbf{69}, 681 (1946).

\bibitem{Goy83} P. Goy, J. M. Raimond, M. Gross, and S. Haroche, Phys. Rev. Lett. \textbf{50}, 1903 (1983).

\bibitem{Kle83} D. Kleppner, Phys. Rev. Lett. \textbf{47}, 233 (1983).
\bibitem{Hul85} R. G. Hulet, R. S. Hilfer, and D. Kleppner, Phys. Rev. Lett. \textbf{55}, 2137 (1985).

\bibitem{Jhe87} W. Jhe, A. Anderson, E. A. Hinds, D. Meschede,
L. Moi, and S. Haroche, Phys. Rev. Lett. \textbf{58}, 666 (1987).

\bibitem{Martinis-86} D. Esteve, M. H. Devoret, and J. M. Martinis, Phys. Rev. B {\bf 34}, 158 (1986).

\bibitem{Blais-04} A. Blais, R. S. Huang, A. Wallraff,
S. M. Girvin, and R. J. Schoelkopf, Phys. Rev. A {\bf 69}, 062320
(2004).

\bibitem{Wal04} A. Wallraff, D. I. Schuster, A. Blais, L. Frunzio,
R. S. Huang, J. Majer, S. Kumar, S. M. Girvin, and R. J. Schoelkopf,
Nature {\bf 431}, 162 (2004).

\bibitem{Hou08} A. A. Houck, J. A. Schreier, B. R. Johnson,
J. M. Chow, Jens Koch, J. M. Gambetta, D. I. Schuster, L. Frunzio,
M. H. Devoret, S. M. Girvin, and R. J. Schoelkopf, Phys. Rev. Lett.
\textbf{101}, 080502 (2008).

\bibitem{Red10} M. D. Reed, B. R. Johnson, A. A. Houck, L. DiCarlo,
J. M. Chow, D. I. Schuster, L. Frunzio, and R. J. Schoelkopf, Appl.
Phys. Lett. \textbf{96}, 203110 (2010).

\bibitem{Sank-14} E. Jeffrey, D. Sank, J. Y. Mutus, T. C. White, J. Kelly, R. Barends, Y. Chen, Z. Chen, B. Chiaro, A. Dunsworth, A. Megrant, P. J. J. O'Malley, C. Neill, P. Roushan, A. Vainsencher, J. Wenner, A. N. Cleland, and J. M. Martinis, arXiv:1401.0257.

\bibitem{Gam11} J. M. Gambetta, A. A. Houck, and A. Blais, Phys. Rev.
Lett. \textbf{106}, 030502 (2011).

\bibitem{Yin13} Y. Yin, Y. Chen, D. Sank, P. J. J. O'Malley,
T. C. White, R. Barends, J. Kelly, E. Lucero, M. Mariantoni, A.
Megrant, C. Neill, A. Vainsencher, J. Wenner, A. N. Korotkov, A. N.
Cleland, and J. M. Martinis, Phys. Rev. Lett. \textbf{110}, 107001
(2013).

\bibitem{Set13} E. A. Sete, A. Galiautdinov, E. Mlinar,
J. M. Martinis, and A. N. Korotkov, Phys. Rev. Lett. \textbf{110},
210501 (2013).

\bibitem{Boi-08} M. Boissonneault, J. M. Gambetta, and A. Blais, Phys. Rev. A {\bf 77}, 060305 (2008).

\bibitem{Boi09} M. Boissonneault, J. M. Gambetta, and A. Blais,
Phys. Rev. A \textbf{79}, 013819 (2009).

\bibitem{Sli12} D. H. Slichter, R. Vijay, S. J. Weber, S. Boutin,
M. Boissonneault, J. M. Gambetta, A. Blais, and I. Siddiqi, Phys.
Rev. Lett. \textbf{109}, 153601 (2012).

\bibitem{Haroche-book} H. Haroche and J.-M Raimond, \textit{Exploring
the Quantum: Atoms, Cavities, and Photons} (Oxford University Press, Oxford,
2006).

\bibitem{Houck-transfer} S. J. Srinivasan, N. M. Sundaresan, D. Sadri,
Y. Liu, J. M. Gambetta, T. Yu, S. M. Girvin, and A. A. Houck, arXiv:1308.3471.


\bibitem{Kor-13} A. N. Korotkov, arXiv:1309.6405, Appendix B.

\bibitem{Pie07} P. Meystre and M. Surgent III, \textit{Elements of
Quantum Optics} (Springer-Verlag, Berlin, 2007).

\bibitem{Car93} H. J. Carmichael, Phys. Rev. Lett. \textbf{70}, 2273 (1993).

\bibitem{Joh12} J. E. Johnson, C. Macklin, D. H. Slichter, R. Vijay, E. B.
Weingarten, J. Clarke, and I. Siddiqi, Phys. Rev. Lett. {\bf 109},
050506 (2012).

\bibitem{Ris12} D. Riste, J. G. van Leeuwen, H.-S. Ku, K. W. Lehnert, and
L. DiCarlo, Phys. Rev. Lett. {\bf 109}, 050507 (2012).

\bibitem{X-mon13} R. Barends, J. Kelly, A. Megrant, D. Sank, E. Jeffrey,
Y. Chen, Y. Yin, B. Chiaro, J. Mutus, C. Neill, P. O'Malley, P.
Roushan, J. Wenner, T. C. White, A. N. Cleland, and John M.
Martinis, Phys. Rev. Lett. \textbf{111}, 080502 (2013).

\bibitem{JC-ladder} Note that the language of the JC ladder of
states in Fig.\ \ref{fig-JC} coresponds to the full Hamiltonian
(\ref{Ham}) rather than the Hamiltonian (\ref{ham}) in the rotating
frame.

\bibitem{Galiautdinov-12} A. Galiautdninov, A. N. Korotkov, and
J. M. Martinis, Phys. Rev. A {\bf 85}, 042321 (2012).

\bibitem{Koch-07} J. Koch, T. M. Yu, J. Gambetta, A. A. Houck,
D. I. Schuster, J. Majer, A. Blais, M. H. Devoret, S. M. Girvin, and
R. J. Schoelkopf, Phys. Rev. A {\bf 76}, 042319 (2007).


\bibitem{Gam08} J. Gambetta, A. Blais, M. Boissonneault, A. A. Houck,
D. I. Schuster, and S. M. Girvin, Phys. Rev. A \textbf{77},
 012112 (2008).

\bibitem{Mol69} B. R. Mollow, Phys. Rev. \textbf{188},1969 (1969).



\end{thebibliography}
\end{document}